\documentclass[11pt,preprint2]{aastex}

\usepackage[utf8x]{inputenc}
\usepackage{rotating}
\usepackage{ragged2e}

\newcommand{\mrsun}{R_\odot}
\newcommand{\mdeg}{^\circ}
\newcommand{\Mm}{{\rm Mm}}

\begin{document}

\title{Energy Input Flux in the Global Quiet-Sun Corona}

\author{Cecilia Mac Cormack\altaffilmark{1}, Alberto M. V\'asquez\altaffilmark{2,1}, Marcelo L\'opez Fuentes and Federico A. Nuevo\altaffilmark{1}}
\affil{Instituto de Astronom\'{\i}a y F\'{\i}sica del Espacio (IAFE), CONICET-UBA,\\ CC 67 - Suc 28, (C1428ZAA) Ciudad Aut\'onoma de Buenos Aires, Argentina.}

\author{Enrico Landi and Richard A. Frazin}
\affil{Department of Climate and Space Sciences and Engineering (CLaSP),\\ University of Michigan, 2455 Hayward Street, Ann Arbor, MI 48109-2143, US.}

\altaffiltext{1}{Departamento de F\'{\i}sica, Facultad de Ciencias Exactas y Naturales (FCEN), Universidad de Buenos Aires (UBA),\\ Pabell\'on I, Ciudad Universitaria, (C1428ZAA) Ciudad Aut\'onoma de Buenos Aires, Argentina.}
\altaffiltext{2}{Departamento de Ciencia y Tecnolog\'{\i}a, Universidad Nacional de Tres de Febrero (UNTREF),\\ Valent\'{\i}n G\'omez 4752, (B1678ABH) Caseros, Provincia de Buenos Aires, Argentina.}

\begin{abstract}
We present first results of a novel technique that provides, for the first time, constraints on the energy input flux at the coronal base ($r\sim\,1.025\,{\rm R}_\odot$) of the quiet-Sun at a global scale. By combining \emph{differential emission measure tomography} (DEMT) of EUV images, with global models of the coronal magnetic field, we estimate the energy input flux at the coronal base that is required to maintain thermodynamically stable structures. The technique is described in detail and first applied to data provided by the \emph{Extreme Ultraviolet Imager} (EUVI) instrument, on board the \emph{Solar TErrestrial RElations Observatory} (STEREO) mission, and the \emph{Atmospheric Imaging Assembly} {(AIA)} instrument, on board the \emph{Solar Dynamics Observatory} (SDO) mission, for two solar rotations with different levels of activity. Our analysis indicates that the typical energy input flux at the coronal base of magnetic loops in the quiet-Sun is in the range $\sim\,0.5-2.0\times 10^5\,{\rm (erg\,sec^{-1}\,{\rm cm}^{-2})}$, depending on the structure size and level of activity. A large fraction of this energy input, or even its totality, could be accounted for by Alfvén waves, as shown by recent independent observational estimates derived from determinations of the non-thermal broadening of spectral lines in the coronal base of quiet-Sun regions. This new tomography product will be useful for validation of coronal heating models in magnetohydrodinamic simulations of the global corona.
\end{abstract}


\section{Introduction} \label{sec:intro}
\justify
The heating of the solar corona remains a main topic of current research in solar physics. While there is a wide consensus in the solar physics community on the magnetic nature of the phenomena responsible for the heating of the million-degree corona, the precise mechanisms by which this occurs are an open field of research and active debate. 

Coronal heating theories are traditionally classified into two broad groups: those based on the dissipation of magnetic stress, informally called DC-heating, and those based on the dissipation of magnetohydrodinamic (MHD) waves, also known as AC-heating. The first group includes models in which the stress produced in the coronal field by photospheric motions is released in situ by reconnection and current sheet formation (see e.g., \citet{parker_1988,priest_2011}). Other MHD simulations \citep{gudiksen_2005} model the system as a numerical enclosure in which the energy is injected by photospheric motions at the base and released by Ohmic dissipation in the corona. Also in the first group are models based on the energy release due to the turbulent interplay between photospheric plasma motions and the magnetic field \citep{gomez_2000}. AC heating mechanisms are based on the propagation of disturbances produced at the feet of the magnetic structures that dissipate at certain atmospheric layers releasing energy that translates into plasma heating \citep{heyvaerts_1983,demoortel_2000,oneill_2005}. Recently, \citet{vanballegooijen_2011} proposed a mixed mechanism by which the diffusion of MHD waves at the chromosphere and transition region (TR) interface produces a turbulent regime that heats the plasma. 

The heating of the plasma in different structures (coronal holes, bright loops in active regions, quiet-Sun corona, etc.) is probably dominated by different physical mechanisms. To advance our understanding of the physics underlying this complex phenomenon, advances on observational constraints are key. The majority of the observational literature on coronal heating focus on active region (AR) structures, where individual bright extreme ultraviolet (EUV) and X-ray loops provide direct diagnostics of magnetic structures \citep{reale_2010,schmelz_2010,aschwanden_2011,klimchuk_2015}. Although not evident from EUV/X-ray images due to its diffuse appearance, the quiet-Sun corona is of course also fully threaded by magnetic fields along which energy transport and deposition takes place. The observational study of the heating in the quiet-Sun diffuse corona has comparatively received less attention \citep{benz_2002,wilhelm_2004,hahn_2014}. 

In general, the aforementioned works provide insight on the heating phenomenon at a local scale, for the specific structures selected for observation, and are affected by line-of-sight projection effects. \emph{Differential emission measure tomography} (DEMT) provides a powerful tool to study the quiet-Sun corona at a global scale and in three dimensions \citep{frazin_2009,vasquez_2016}. Based on full solar rotation time series of EUV images taken in channels sensitive to different temperatures, DEMT provides three-dimensional (3D) maps of the electron density and temperature of the lower corona, in the heliocentric height {range 1.02 to 1.225}\,${\rm R}_\odot$. The coronal magnetic field in these regions can be globally modeled by means of \emph{potential field source surface} (PFSS), or MHD models. Combination of the DEMT and global magnetic models has provided useful insight in the 3D thermodynamical structure of the global quiet-Sun corona \citep{huang_2012,nuevo_2013,nuevo_2015}, making it an ideal tool to provide constraints on coronal heating for these regions. However, until now such a DEMT application had not been developed. 

In this work, we develop the first version of a new DEMT tool capable of providing constraints on the coronal heating of the quiet-Sun global corona. Specifically, it provides spatial two-dimensional (2D) maps of the energy input flux required at the coronal base to maintain thermodynamically stable coronal structures under hydrostatic assumption. This new tool is applied to two specific selected rotations with different levels of activity, studied by means of two EUV instruments, namely the \emph{Extreme Ultraviolet Imager} (EUVI, \citealt{wuelser_2004}), on board the \emph{Solar TErrestrial RElations Observatory} (STEREO) mission, and the \emph{Atmospheric Imaging Assembly} (AIA, \citealt{lemen_2012}) telescope on board the \emph{Solar Dynamics Observatory} (SDO) mission.

Section \ref{data} summarizes the techniques, instruments and data sets used. Section \ref{model} details the energy model and the loop-integrated quantities that are introduced, which form the new DEMT tool. Section \ref{demt-pfss_results} shows the DEMT results for the selected periods, while Section \ref{phis_results} details the new results on energy input flux at the coronal base. In Section \ref{ebtel} these results are compared to a 0D hydrodynamic (HD) model. Section \ref{conc} discusses the main conclusions and its implications, and anticipates further planned work.

\section{Methodology and Data} \label{data}

This work is based on the study of the global corona by means of the DEMT technique to determine its 3D thermodynamical structure, the PFSS modeling of its global magnetic field, and the combination of both. The technique is applied to two specific Carrington rotations (CRs): CR-2081 (2009, 09 March through 05 April), a deep minimum period between solar cycles (SCs) 23 and 24 characterized by virtually no ARs, and CR-2099 (2010, 13 July through 9 August), a rotation during the early rising phase of SC 24. 

{Both periods were tomographically reconstructed from data taken by the EUVI/STEREO instrument. In the case of CR-2099, the period was also reconstructed using data taken by the AIA/SDO instrument. For both rotations, the magnetic field was modeled by means of the \emph{Finite Difference Potential-Field Solver} (FDIPS) {PFSS} model developed by \citet{toth_2011}, using as boundary conditions synoptic magnetograms built from data taken by the \emph{Michelson Doppler Imager} (MDI, \citealt{scherrer_1995}) on board the Solar and Heliospheric Observatory (SOHO) mission.}

{In DEMT the inner corona, in the height range $1.0-1.25\,\mrsun$, is discretized in a spherical computational grid. The size of the grid cell (or voxel) is set to $0.01\,\mrsun$ in the radial direction (or $\sim 7\,\Mm$), and to $2\,\mdeg$ in both the latitudinal and longitudinal directions (or $\sim 27\,\Mm$ at an intermediate grid height of $r=1.1\,\mrsun$ at the equator). With this angular resolution one image every 6 hours is the cadence needed to fully constrain the inversion problem, for a total of about 110 images to cover a full solar rotation. This is the standard DEMT resolution used over the past few years in all previously published work based on this technique \citep{vasquez_2016}, providing a good compromise between resolution and computational load.}

Two consecutive procedures are then performed. In a first step, time series of EUV images in different bands, covering a full solar rotation, are used to perform EUV tomography. The product of the tomographic inversion in each band is the 3D distribution of the \emph{filter band emissivity} (FBE), defined as the wavelength integral of the coronal EUV spectral emissivity and the telescope's passband function of each band. 

In a second step, the FBE values obtained for all bands in each tomographic cell (or voxel) are used to constrain the determination of a \emph{local differential emission measure} (LDEM) distribution. {The LDEM describes the temperature distribution of the electron plasma contained in each individual tomographic grid voxel. The LDEM is defined so that the electron density $N_e$ and electron mean temperature $T_m$ (averaged over temperature) of each voxel are computed as,}

\begin{eqnarray}
 N_e^2 &=& \int\,dT\,{\rm LDEM}(T),\label{Ne} \\
 T_m   &=& \frac{1}{N_e^2}\,\int\,dT\,T\,{\rm LDEM}(T).\label{Tm}
\end{eqnarray}

{Due to their ill-posed nature DEM inversion problems are difficult to treat, both in bright loops observed in ARs, as well as in the diffuse quiet Sun corona here analyzed. If based on high resolution spectral data, DEM analysis is best performed through the Monte Carlo Markov Chain (MCMC) approach \citep{kashyap_1998}, or regularized inversion techniques \citep{hannah_2012} (also applicable to filter band data sets). In the case of narrowband filter telescopes, the AIA instrument has improved temperature diagnostic capabilities over previous EUV telescopes instruments (such as EUVI). Application of MCMC methods based on AIA images have been explored by several works \citep{testa_2012, delzanna_2013}, as discussed in \citet{nuevo_2015}. \citet{schmelz_2013} have recently shown advances of MCMC multitermal DEM analysis based on AIA data of bright EUV loops in ARs, using all six AIA channels simultaneously, with updated temperature response functions based on CHIANTI v7.1. (see also \citealt{schmelz_2016} for an application to an extended study of bright loops).} 

{In the case of DEMT, the DEM parametric approach is suitable due to the limited number of data points. \citet{nuevo_2015} have developed a detailed study of the capabilities of the parametric technique in DEMT when applied to both EUVI and AIA data. In the case of EUVI data its three coronal bands are used (171, 195, and 284 \AA), with maximum sensitivity temperatures in the range 1.0-2.15 MK (see Table 1 in \citealt{nuevo_2015}). In the case of AIA data, as DEMT targets the diffuse quiet Sun corona, the filters that are currently used in DEMT are those of 171, 193, 211, and 335 \AA, with maximum sensitivity temperatures in the range 0.85-2.5 MK (see Table 1 in \citealt{nuevo_2015}). These four AIA bands cover the main temperature range characteristic of the quiet Sun regions targeted by DEMT. Adding the 94 and 131 \AA\ bands of AIA in DEMT studies has also been attempted (exploring other parametric models for the LDEM as well), but at a global scale (as required in tomography) the signal from these channels in the diffuse quiet Sun is in general too weak to provide meaningful information.} 

{In this study the EUVI and AIA temperature responses have been computed based on CHIANTI v7.1, as discussed in detail in \citet{nuevo_2015}. As shown in that work, when using the aforementioned four coronal bands of AIA, the tomographic emissivity 3D distributions are best explained by an ubiquitous bimodal LDEM, described by two Gaussian distributions with characteristic centroids of order 1.5 MK and 2.6 MK, dubbed ``warm'' and ``hot", respectively. {In the case of using all three EUVI channels DEMT detects only the warm component. It is shown 1) that if only the three AIA channels of 171, 193 and 211 \AA\ are used (as in this work) then also only the warm component is detected, and 2) that in that case the results are very similar to those based in EUVI data, with some small systematic differences due to the different response functions of the respective trios of filters. The LDEM obtained are typically broad, with variable FWHM depending on the region. Indeed, in that same work, a controlled study inverting synthetic data from modeled DEM distributions has shown DEMT to be able to detect nearly isothermal and multithermal plasmas, reflected in the resulting value of the FWHM of the LDEM.}}

{The parametric approach is at the moment the best implementation of DEMT. Even with its limitations, the resulting LDEM distributions, as well as the electron density and mean temperature maps they produce, have been validated with other types of observational studies. {For example, in \citet{nuevo_2015} the LDEM parametric technique has been validated comparing its results when applied to LOS-DEM analysis of image pixels against those obtained by other authors using MCMC techniques (based on EIS data). Similar results are obtained, in particular in terms of the bimodal DEMs. Applied to many different rotations, DEMT has so far provided consistent results across many studies and coronal regions, which have also been used as a validation tool for MHD simulations of the global corona (see review in \citealt{vasquez_2016})}. Future new developments could explore the implementation of regularized inversion techniques for the determination of the LDEM at each voxel, or even MCMC methods as applied to AIA images by \citet{schmelz_2016}, who have been able to exploit those if at least 3 data points are available.}

{In this work, all three coronal bands of EUVI are used to study CR-2081 and CR-2099, and in the latter case an alternate study based on the AIA trio 171-193-211 \AA\ is also carried out. When using three bands, as in this work,} the LDEM is modeled by a Gaussian function \citep{nuevo_2015} dependent on three free parameters: centroid temperature, temperature width, and total area. In each voxel the values of the free parameters are found so that the synthetized emissivity values ${\rm FBE}^{(k)}_{\rm syn}$ best match the tomographic values ${\rm FBE}^{(k)}_{\rm tom}$ for all 3 bands $k=1,2,3$, achieved by minimizing the score,

\begin{equation}\label{Rdef}
R = \frac{1}{3}\sum_{k=1}^{3}
 \left| \, 1 - {\rm FBE}^{(k)}_{\rm syn} / \,{\rm FBE}^{(k)}_{\rm tom}\, \right|.
\end{equation}

\noindent
{A score of $R\sim 0$ means the LDEM accurately predicts the 3 tomographic FBEs at a given voxel, while higher scores mean a less accurate prediction.}

{Once the LDEM is obtained in each voxel of the tomographic computational grid, Equations (\ref{Ne}) through (\ref{Rdef}) allow computation of 3D maps of the electron density $N_e$, electron mean temperature $T_m$, and score $R$.  The reader is referred to \citet{frazin_2009} and \citet{nuevo_2015} for a detailed description of the DEMT technique, and to \citet{vasquez_2016} for a recent review on all published work based on it.}

The DEMT results can be then combined with the global magnetic field model, by tracing the results of the former along the magnetic field lines of the latter. {This approach was first used by \citet{huang_2012} to study the temperature structure of the solar corona during the last minimum, and later on applied by \citet{nuevo_2013} to expand the analysis to rotations with different level of activity.} In this paper {the same approach is used to study}, in an original fashion, the energy balance along individual magnetic loops, {as it is described} in Section \ref{model}. In doing so, {a new set of numerical tools within the suite of DEMT codes was developed}.

\section{Loop Model and Energy Balance}\label{model}

\begin{figure*}
\centering
\includegraphics[scale=0.4]{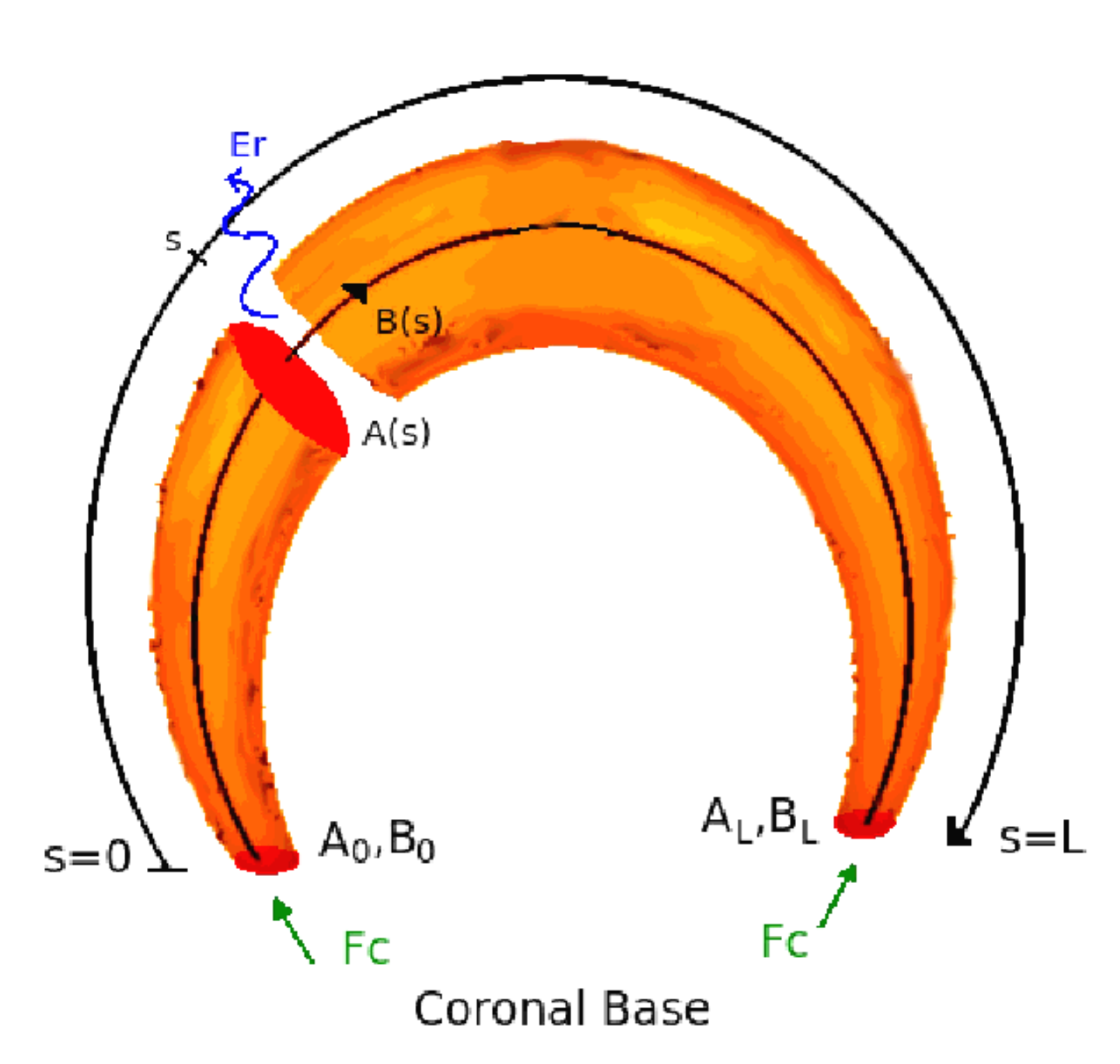}
\caption{A closed coronal magnetic flux tube and the coordinate axis $s$ along it. The radiative power (per unit volume) $E_r(s)\,[{\rm erg\,sec^{-1}\,cm^{-3}}]$ emitted at an arbitrary position $s$ along the tube is indicated, along with the local transverse area $A(s)$ of the tube. The conductive heat flux $F_c$ at both footpoints of the magnetic tube are also sketched.}
\label{fig_01}
\end{figure*}

A simple hydrostatic model for a coronal magnetic flux tube, sketched in Figure \ref{fig_01} is considered. The position along the tube is given by the variable $s$, with $s=0$ and $s=L$ representing the positions at the coronal base. At any position $s$, the (unknown) coronal heating power $E_h(s)$ is balanced by the two major coronal losses, namely the radiative power $E_r(s)$ and the thermal conduction power $E_c(s)$ \citep{aschwanden_2004},

\begin{equation}\label{Balance}
E_h(s) = E_r(s)+ E_c(s),
\end{equation}

\noindent
where the three power quantities are per unit volume, i.e., have units of $[{\rm erg\,sec^{-1}\,cm^{-3}}]$. In the optically thin corona the radiative energy is emitted isotropically, while the heat conductive flux is constrained to flow along magnetic field lines. Considering the coronal magnetic flux tube as a whole, the conductive flux at the coronal base represents a net gain or loss of energy for the system. 

The thermal conduction power $E_c$ equals the divergence of the conductive heat flux $F_c$, which can be expressed as the derivative along the flux tube,

\begin{equation}\label{Ec}
E_c(s)=\frac{1}{A(s)}\,\frac{d}{ds}\left[A(s)\,F_c(s)\right].
\end{equation}

\noindent 
where, for the quiescent solar coronal plasma regime, the conductive heat is dominated by the electron thermal conduction described by the usual Spitzer model \citep{spitzer_1962},

\begin{equation}\label{Fc}
F_c(s)=-\kappa_0\,{T(s)}^{5/2}\,\frac{dT}{ds}(s),
\end{equation}

\noindent 
where the Spitzer thermal conductivity is $\kappa_0 \approx 9.2\times 10^{-7}\,{\rm erg}\,{\rm s^{-1}}\,{\rm cm^{-1}}\,{\rm K^{-7/2}}$.

For each of the three power quantities per unite volume in Equation (\ref{Balance}), a corresponding total power $\gamma\,[{\rm erg\,sec^{-1}}]$ in the coronal part of the magnetic flux tube is obtained by integrating them over its whole volume,

\begin{equation}\label{gammas}
\gamma_{i} \equiv \int_0^L ds\,A(s)\,E_{i}(s). 
\end{equation}

\noindent
where $i=c,r,h$ denotes the conductive, radiative, and heating terms, respectively. Using Equation (\ref{Ec}), the integrated thermal conduction power can be expressed as,

\begin{equation}\label{gammac}
\gamma_c = A_L\,F_{c,L} - A_0\,F_{c,0},
\end{equation}

\noindent
where $A_0$ and $A_L$ are the values of the transversal area at $s=0$ and $s=L$, respectively, and $F_{c,0}$ and $F_{c,L}$ are the respective values of the conductive heat flux.

Dividing the three integrated power quantities in Equation (\ref{gammas}) by the total basal area of the flux tube {$A_0+A_L$, three associated} flux quantities $\phi\,[{\rm erg\,sec^{-1}\,{\rm cm}^{-2}}]$ can be defined as,

\begin{equation}\label{phis}
\phi_{i}\equiv\frac{\gamma_{i}}{A_0+A_L}; \,\,\,i=h,r,c.
\end{equation}

Using these last three quantities, integration of Equation (\ref{Balance}) over the whole coronal volume of the magnetic flux tube implies the integrated energy balance,

\begin{equation}\label{FluxBalance}
\phi_h = \phi_r + \phi_c.
\end{equation}

{The energy required to heat the plasma contained in the magnetic flux tube is ultimately injected into it through its coronal base. The quantity $\phi_h$ represents the total energy input flux due to all mechanisms except heat conduction (accounted for by the term $\phi_c$). The quantity $\phi_h$ will hereafter be referred to as the ``energy input flux" at the coronal base. The quantity $\phi_c$ is total energy flux entering (or leaving) the coronal part of the magnetic tube due to heat conduction. The quantity $\phi_r$ is the total radiative power emitted by the plasma contained in the coronal part of the magnetic flux tube, divided by the total area of its coronal base. Begging proportional to the squared local electron density, which decreases strongly with height, most of the the radiative loss occurs in the lower heights, so that $\phi_r$ provides a characteristic value of the coronal radiative flux.}

The magnetic null divergence condition, integrated along the magnetic flux tube, reads $A(s)\,B(s)=A_0\,B_0=A_L\,B_L{=c}$, where $c$ is a constant specific to each magnetic flux tube, and $B_0$ and $B_L$ are the values of the magnetic field strength at $s=0$ and $s=L$, respectively. Using these relations into Equations (\ref{gammas}), (\ref{gammac}), and (\ref{phis}), the radiative and conductive terms of the RHS of Equation (\ref{FluxBalance}) can be expressed as

\begin{eqnarray}
\phi_r &=& \left(\frac{B_0\,B_L}{B_0+B_L}\right)\,\int_{0}^L ds\,\frac{E_r(s)}{B(s)}\label{phir2},\label{phir2}\\
\phi_c &=& \frac{B_0\,F_{c,L}-B_L\,F_{c,0}}{B_0+B_L} \label{phic}.\label{phic2}
\end{eqnarray}

Note that, by introducing these integrated flux quantities, the energy balance equation (\ref{FluxBalance}) is freed from transversal area values, holding then for individual magnetic \emph{field lines}, rather than magnetic flux tubes. {In studies combining DEMT with magnetic models, individual magnetic field lines from the model are assigned the DEMT values of $N_e$ and $T_m$ of the tomographic voxels through which they pass through (as described in Section \ref{trace} below). The DEMT values for density and temperature in each computational voxel represent an average description of the plasma contained in it. Assigning these values to field lines of the magnetic model provides then a semi-empirical steady state model for quiet Sun long-lived coronal structures, aiming at describing their average state over the DEMT temporal resolution ($\sim 1/2$ solar rotation). This approach has been previously used by \citet{huang_2012} and \citet{nuevo_2013} to study thermodynamical properties of the large scale quiet Sun corona, and also served as validation tool for steady-state MHD 3D models of the global corona \citep{jin_2012,evans_2012,oran_2015}. It is out of the capabilities of DEMT to analyze individual bright loops seen in ARs, or the fast dynamics of any structure. This study, and in particular the balance Equation \ref{FluxBalance}, is to be understood also as an average description of the energy balance in each loop.}

The magnetic field strength $B(s)$ along the field line and, in particular, its values at the coronal base $B_0$ and $B_L$, are one of the products directly available from the global corona magnetic extrapolation. Hence, the terms involved in Equations (\ref{phir2})-(\ref{phic2}) can be computed from the results of DEMT and the magnetic extrapolation. 

In the optically thin corona, the radiative power of an isothermal plasma at temperature $T$ is computed as $E_r = N_e^2\,\Lambda(T)$, where the radiative loss function $\Lambda(T)$ is in turn computed by means of a model, such as the CHIANTI atomic database and plasma emission model \citep{dere_1997}, used in this work in its latest version. {Hence, at each tomographic voxel the radiative power is computed from the temperature distribution ${\rm LDEM}(T)$ obtained from the DEMT technique as,}

\begin{equation}\label{Er}
E_r = \int\,dT\,{\rm LDEM}(T)\,\Lambda(T).
\end{equation}

\noindent
where, from Equation (\ref{Ne}), $dT\,{\rm LDEM}(T)$ is the contribution to the quadratic electron density in the voxel of the plasma with temperature in the range $T\pm dT$. {The radiative power $E_r$, in itself a novel DEMT product introduced in this work, can be numerically traced along the field lines of the global magnetic coronal extrapolation, as explained in Section \ref{trace} below. This allows computation of the quantity $\phi_r$ from Equation (\ref{phir2}) for each magnetic field line in the model.}

Finally, the quantity $\phi_c$ given by Equation (\ref{phic2}) requires computation of the conductive heat flux $F_c$ at both coronal base points of the field line. To this end, in Equation (\ref{Fc}) the temperature $T$ and temperature gradient $dT/ds$ are computed from the DEMT results traced along the each magnetic field line.

The next section details the numerical implementation of the tracing of the DEMT results along the magnetic field lines, and the computation of the corresponding quantities $\phi_r$ and $\phi_c$. Once these two quantities are computed for each field line, Equation (\ref{FluxBalance}) {allows computation of the energy input flux at the coronal base $\phi_h$, the new DEMT product that constitutes the main result of this work.}

\section{DEMT-PFSS Results}\label{demt-pfss_results}

\subsection{3D Reconstruction of Density and Temperature}\label{demt_results}

{This section shows and describes} the DEMT results corresponding to the two rotations, CR-2081 and CR-2099, based on EUVI data in both cases, as well as AIA/SDO data in the latter one. In the case of CR-2099, the very similar EUVI based results {are omitted to save space}. CR-2099 has been the subject of DEMT analysis based on both EUVI and AIA data in \citet{nuevo_2015}, where the results are compared in detail {and shown to be consistent. Density values obtained with AIA are $\sim2\%$ smaller compared those obtained with EUVI, while temperatures are $\sim 8\%$ larger. These systematic differences are due to slight differences in the temperature response between respective channels of both instrumental sets.}

The 3D distribution of the FBE was tomographically reconstructed for the EUVI bands of 171, 195 and 284 \AA\, in both selected rotations, as well as for the AIA bands of 171, 193 and 211 \AA\, in the case of CR-2099. Using the three FBE reconstructions in each rotation, the 3D distribution of the LDEM was found. {Finally, from Equations (\ref{Ne}) through (\ref{Rdef}), the 3D maps of the electron density $N_e$, electron mean temperature $T_m$, and score $R$ were computed for both rotations.}

\begin{figure*} 
\begin{center}
\begin{minipage}{0.48\textwidth}
\includegraphics[width=\textwidth]{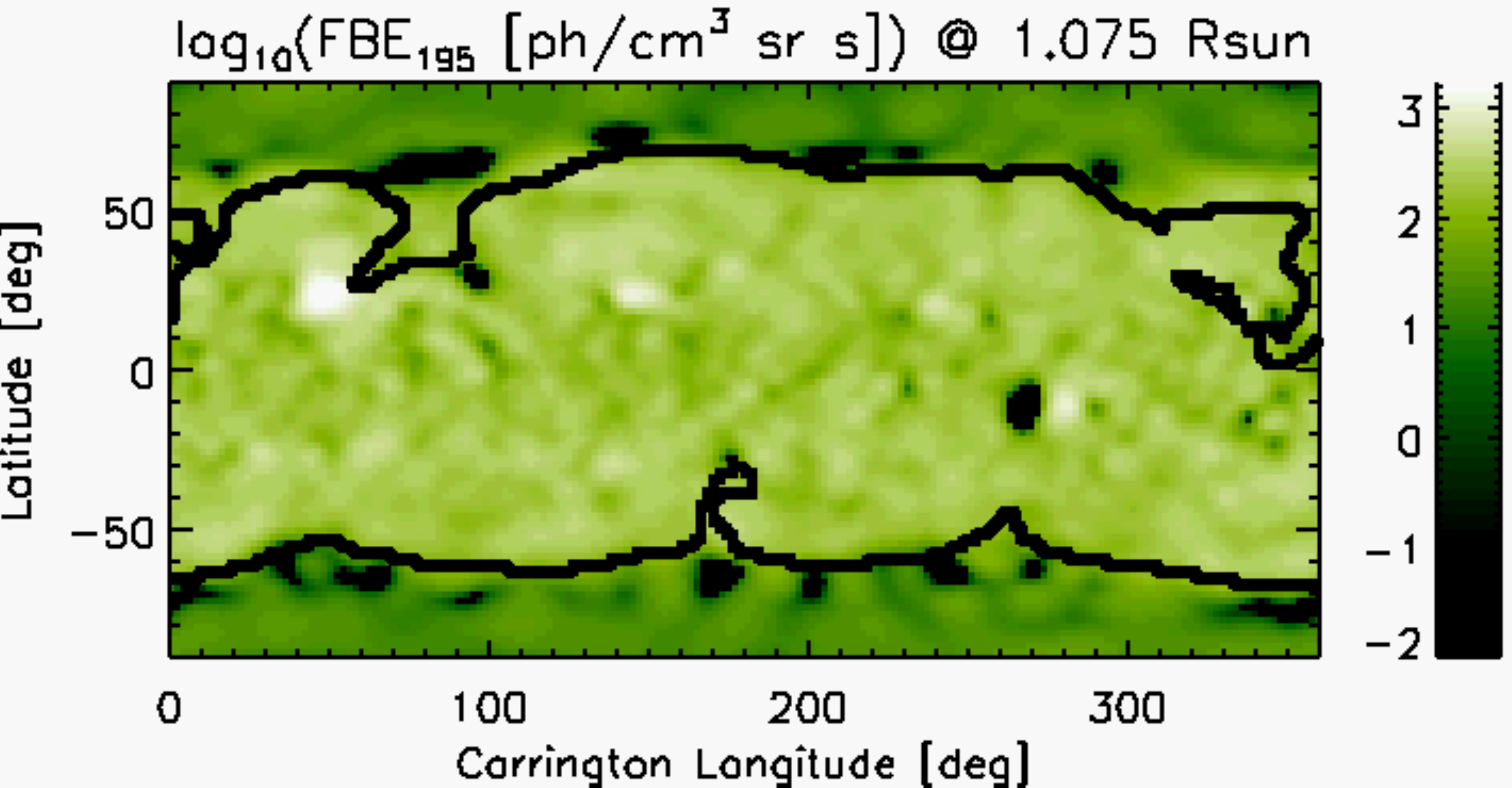}
\includegraphics[width=\textwidth]{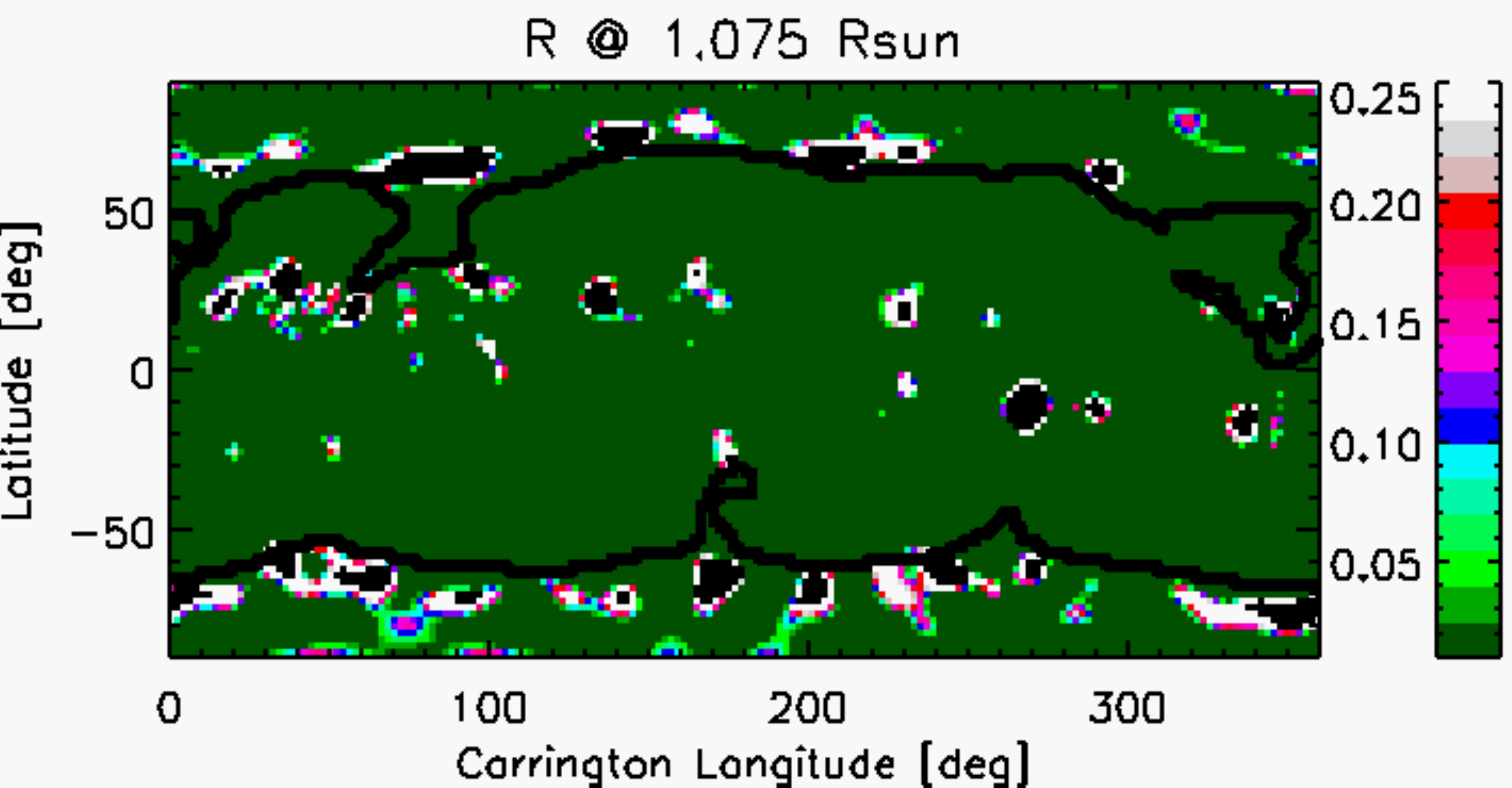}
\includegraphics[width=\textwidth]{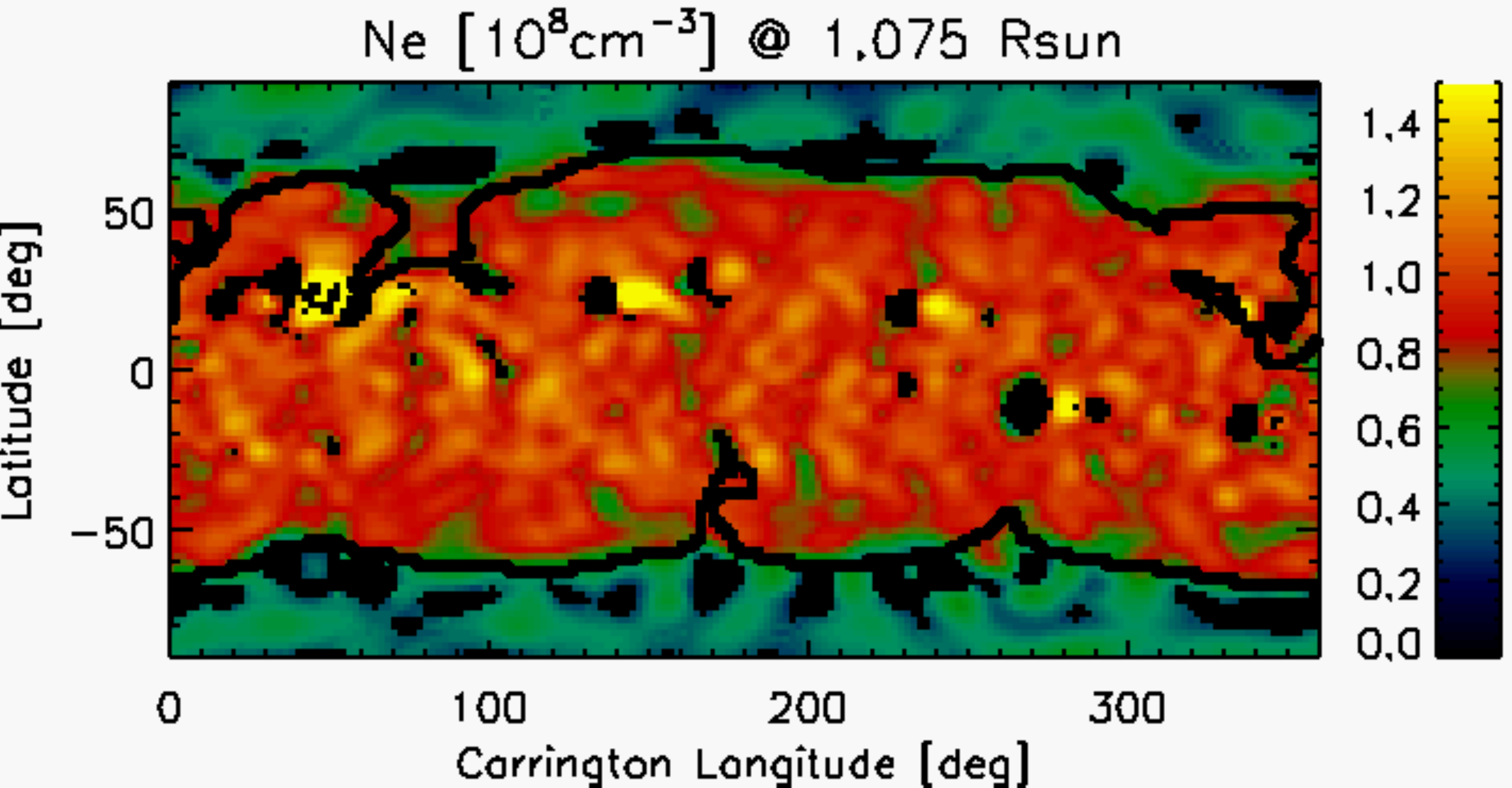}
\includegraphics[width=\textwidth]{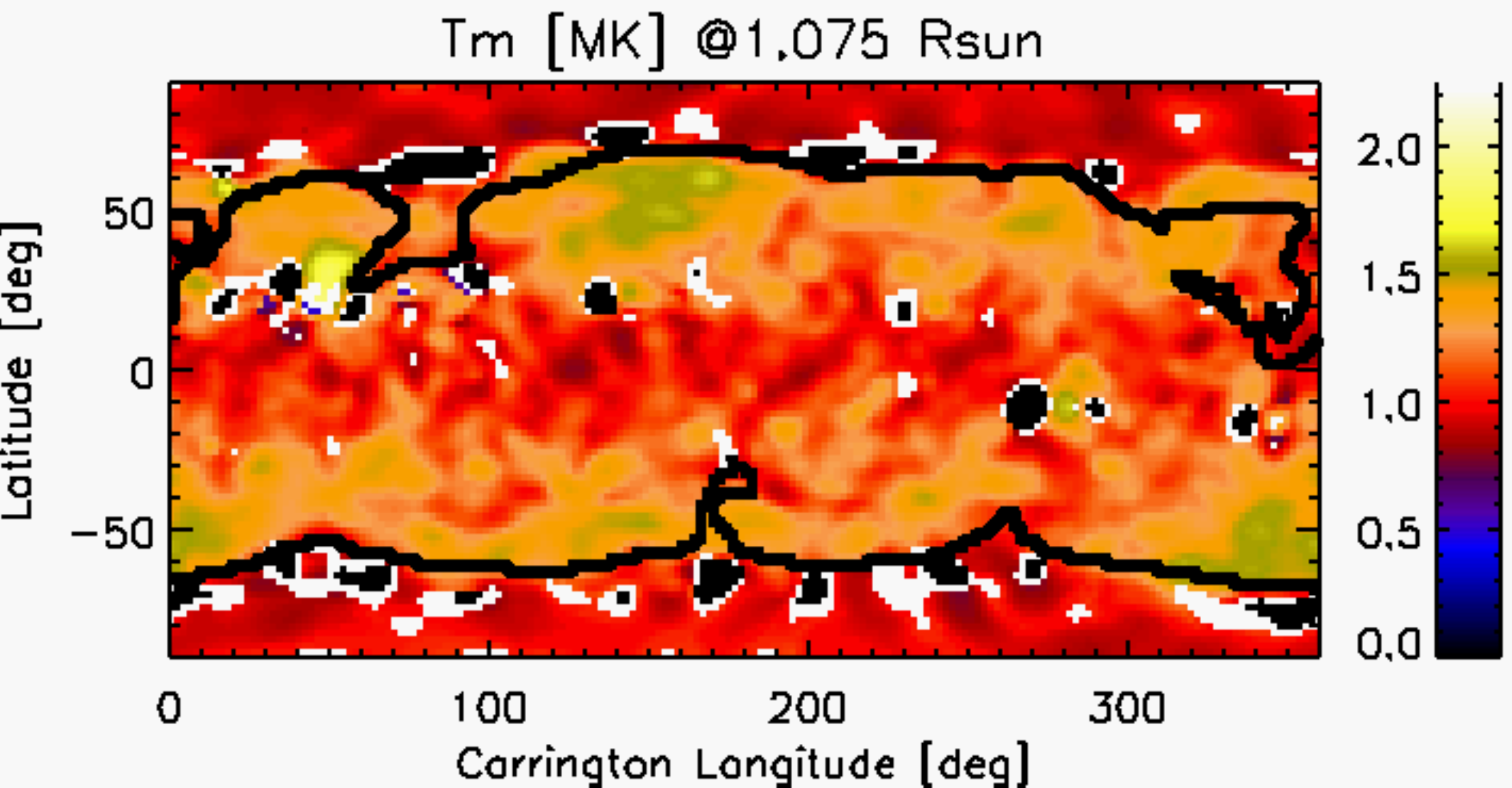}
\end{minipage}
\begin{minipage}{0.48\textwidth}
\includegraphics[width=\textwidth]{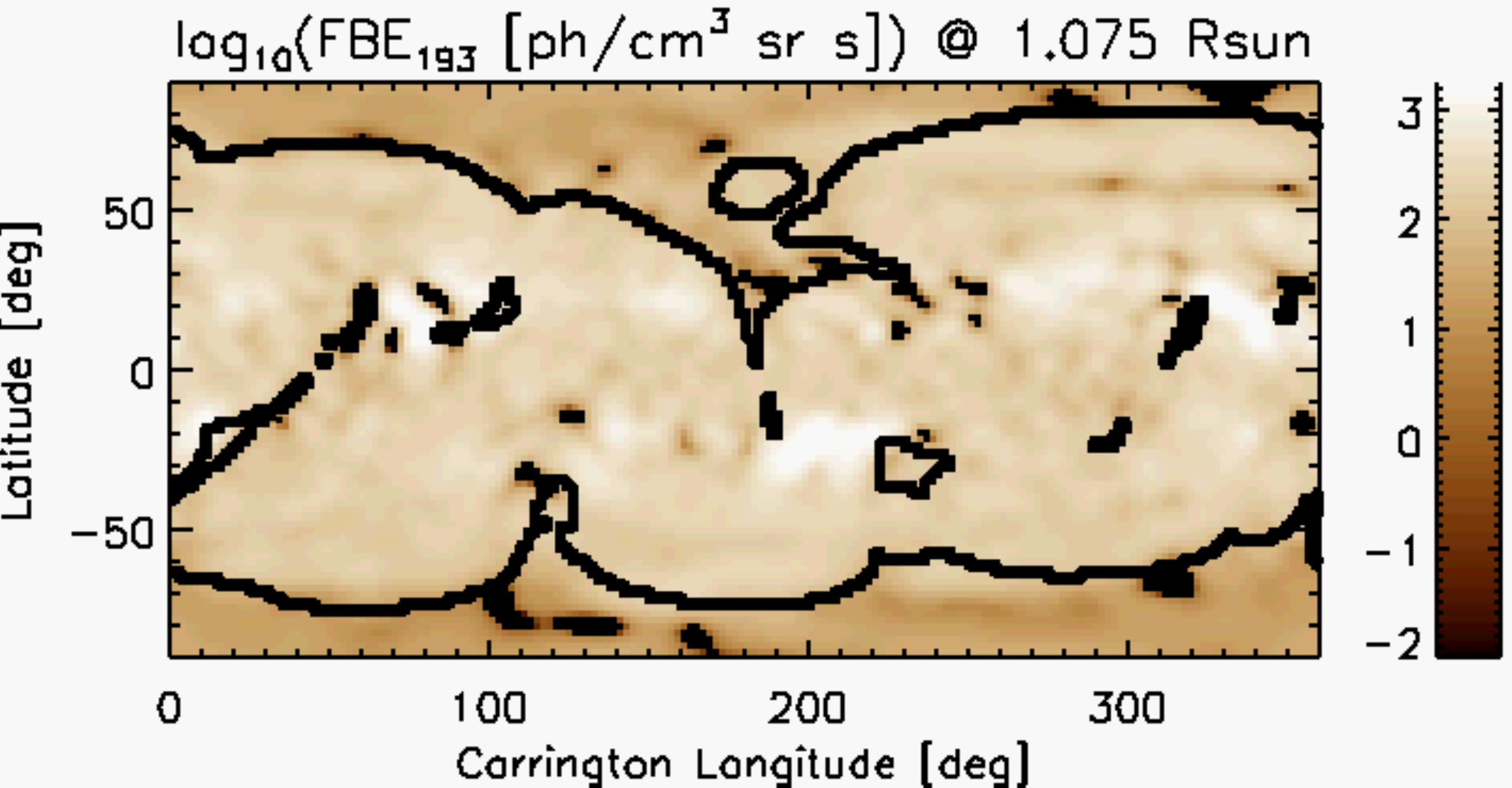}
\includegraphics[width=\textwidth]{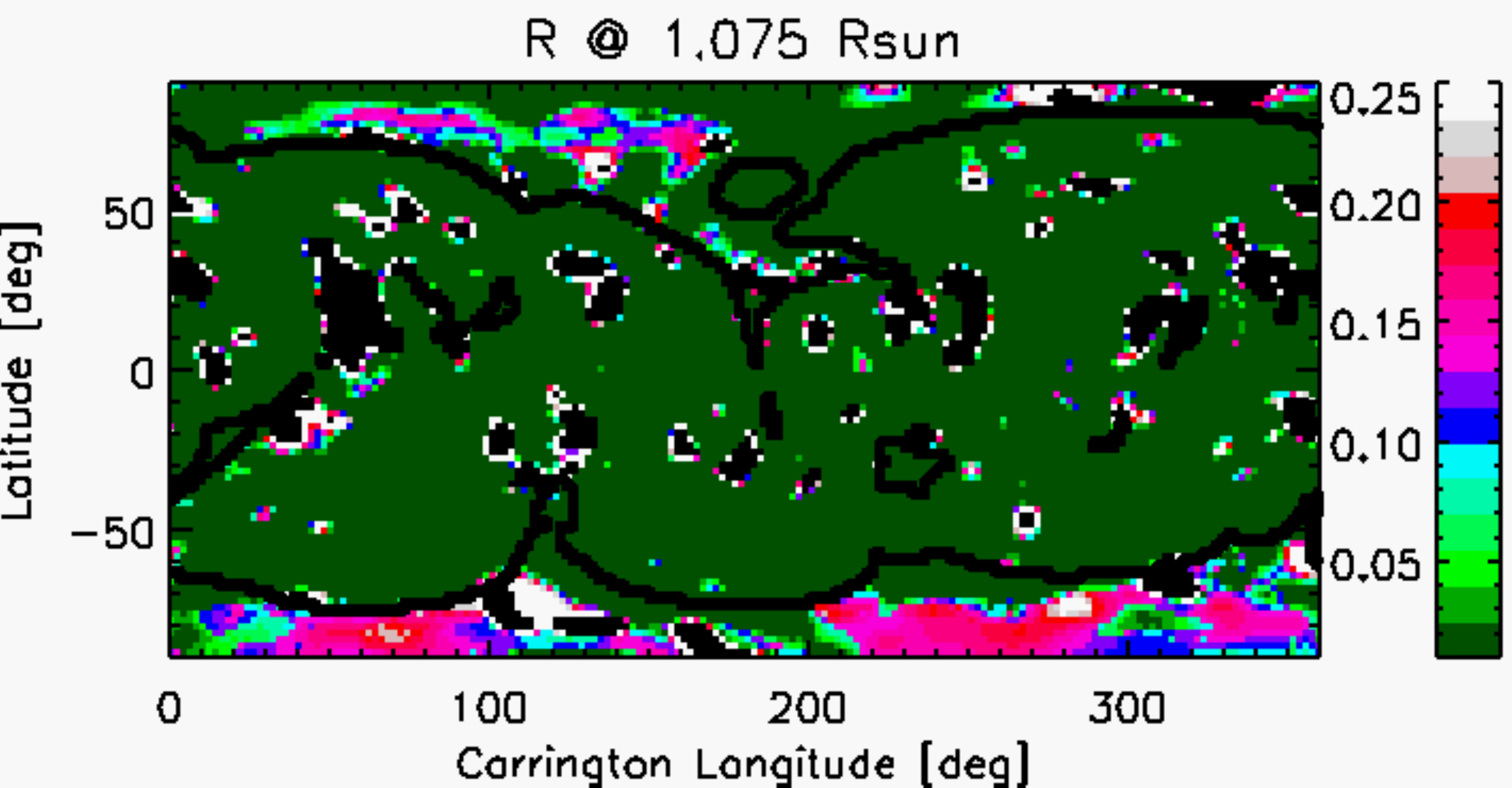}
\includegraphics[width=\textwidth]{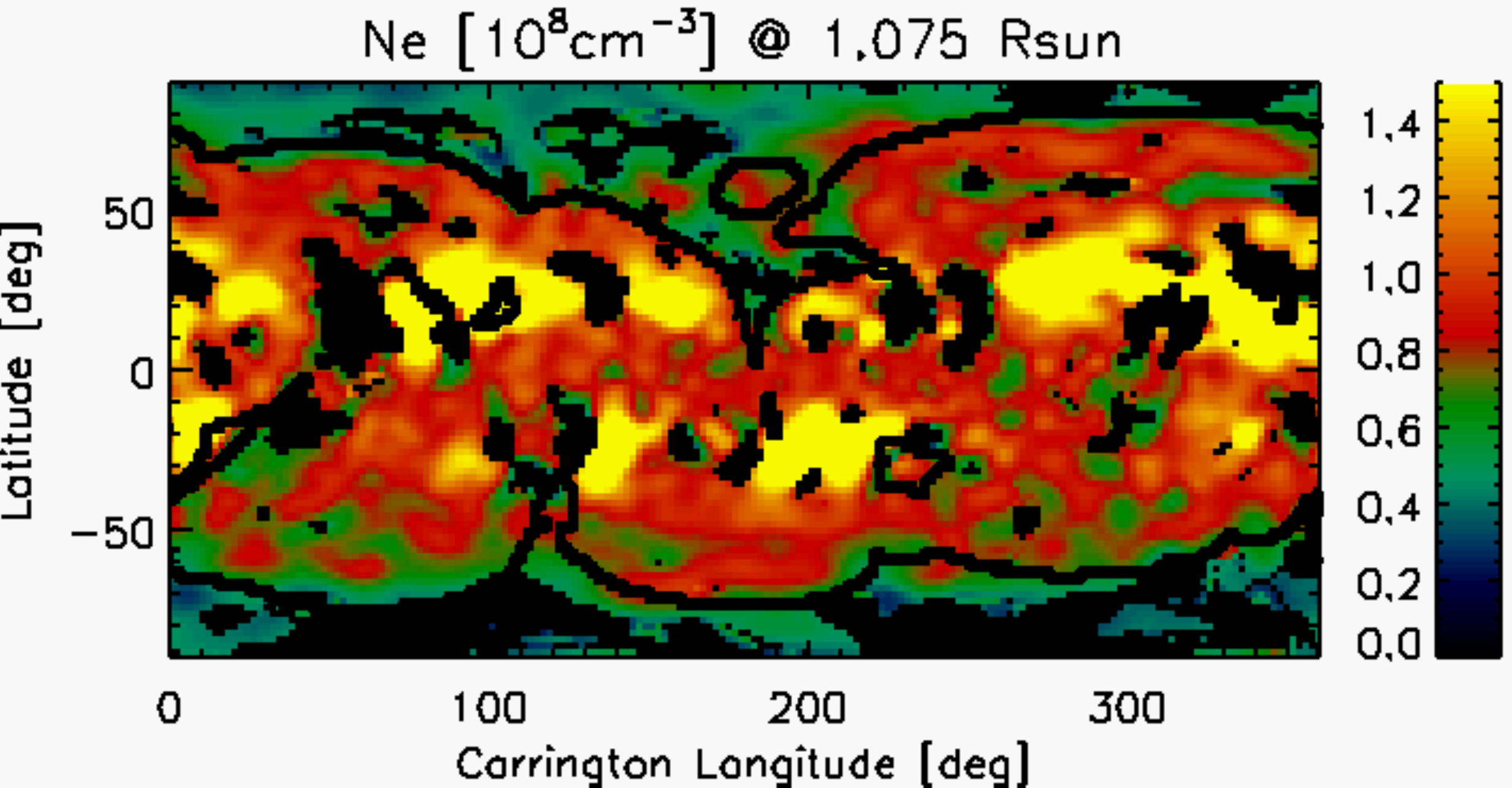}
\includegraphics[width=\textwidth]{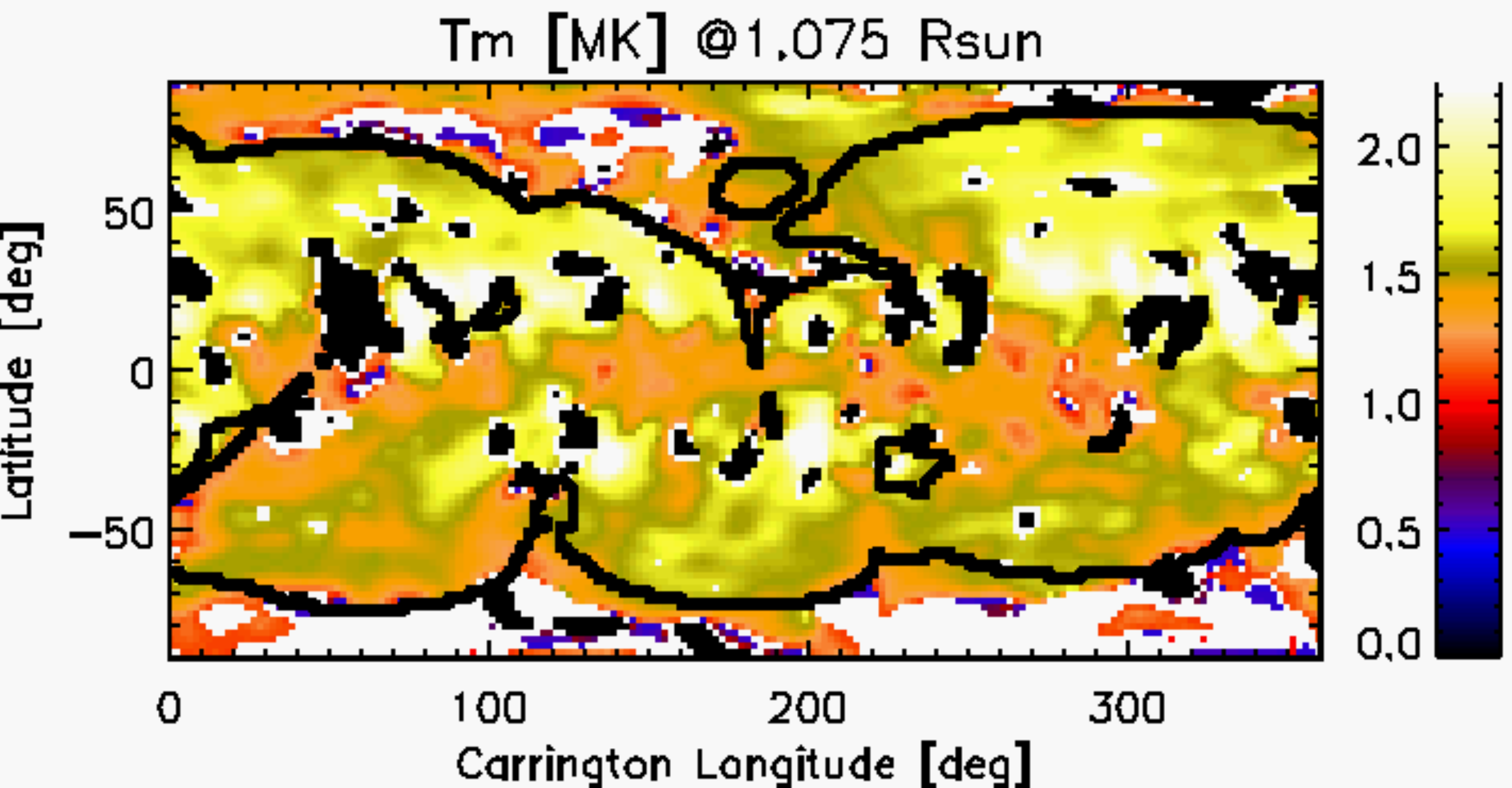}
\end{minipage}
\end{center}
\caption{DEMT results at $r=1.075$\,R{$_\odot$} for CR-2081 based on EUVI data (left panels) and for CR-2099 based on AIA data (right panels). The top panels show Carrington maps of the reconstructed FBE $[{\rm ph\,cm^{-3}\,sr^{-1}\,s^{-1}}]$ in the bands EUVI 195 \AA\ (left) and AIA 193 \AA\ (right). The second to fourth row panels show, in descending order, Carrington maps of the following DEMT results: $R$-score (see text), electron density $N_e\,[{\rm 10^8\,cm^{-3}}]$ and electron mean temperature $T_m\,[{\rm MK}]$. In all panels, the overplotted thick-black curves indicate the boundary between the open and closed magnetic regions, as derived from the PFSS model.}
\label{resultsDEMT}
\end{figure*}

As way of example, Figure \ref{resultsDEMT} shows the results obtained from the DEMT analysis at a selected height of the tomographic computational sphere. {The results are shown as \emph{Carrington maps}, the distribution of quantities in latitude and longitude.} Similar maps are obtained at all heights of the tomographic grid. The left panels correspond to the EUVI data based CR-2081 reconstruction, while the right panels do to the AIA data based CR-2099 reconstruction.

The top panels show, as an example, FBE maps in the bands of EUVI 195 \AA\, and AIA 193 \AA. {Due to unresolved coronal dynamics, tomographic reconstructions exhibit artifacts such as smearing and negative values of the reconstructed FBEs, or zero when the solution is constrained to positive values. These are called zero-density artifacts (ZDAs). In the FBE Carrington maps, ZDA voxels are indicated as {solid-black regions. The overplotted thick-black curves in all panels indicate the boundary between the open and closed magnetic regions, as derived from the PFSS model.}

The second row of panels in Figure \ref{resultsDEMT} shows Carrington maps of the $R$, {with ZDA voxels indicated as solid-black regions.}  It can be readily seen that, in both cases, most of the closed-corona volume is characterized by $R<10^{-2}$ (dark-green color), meaning that the LDEM predicts the tomographic FBEs with a precision better than 1\%. Note that, in the case of the AIA analysis, some regions of the open corona are characterized by somewhat larger $R$ scores, but in any case the focus of this paper is the closed corona. 

The bottom two rows of panels show the DEMT electron density $N_e$ and electron mean temperature $T_m$. In the temperature maps, solid-black regions indicate ZDA voxels, while white solid-white regions indicate voxels for which the parametric LDEM has a score $R > 0.1$, i.e. the discrepancy between the tomographic and synthetic FBEs is more than 10\%. In these cells, dubbed \emph{anomalous emissivity voxels} (AEVs), the Gaussian LDEM model does not accurately reproduce the tomographic results. In the electron density maps, ZDAs and AEVs are indicated as black regions.

It is interesting to note that the open/closed {boundaries of the PFSS model quite accurately match contour levels} of the tomographic density and temperature. In other words, along the open/closed boundary the gradient of the tomographic results is approximately perpendicular to it. This is an interesting consistency check between the PFSS and DEMT models, which can also be verified in all previous DEMT works.

{In the tomographic density maps, the voxels with the largest density values (yellowish regions) always correspond to observed active regions (AR). This has been verified by careful comparison of the reconstructions to the catalogue provided by the National Oceanic and Atmospheric Administration (NOAA) Space Weather Prediction Center.\footnote{\tt www.swpc.noaa.gov/ftpmenu/warehouse/2010.html} These regions are characterized by threshold values of the tomographic density, as detailed in \citet{nuevo_2015}. As tomographic reconstructions are not suitable to study the fast-evolving ARs, these regions are left out of the analysis of this paper. Voxels belonging to ZDA and AEV regions are also excluded from the analysis.}

\subsection{Tracing of Results Along Magnetic Field Lines}\label{trace}

\begin{figure*} 
\begin{center}
\begin{minipage}{0.48\textwidth}
\includegraphics[width=\textwidth]{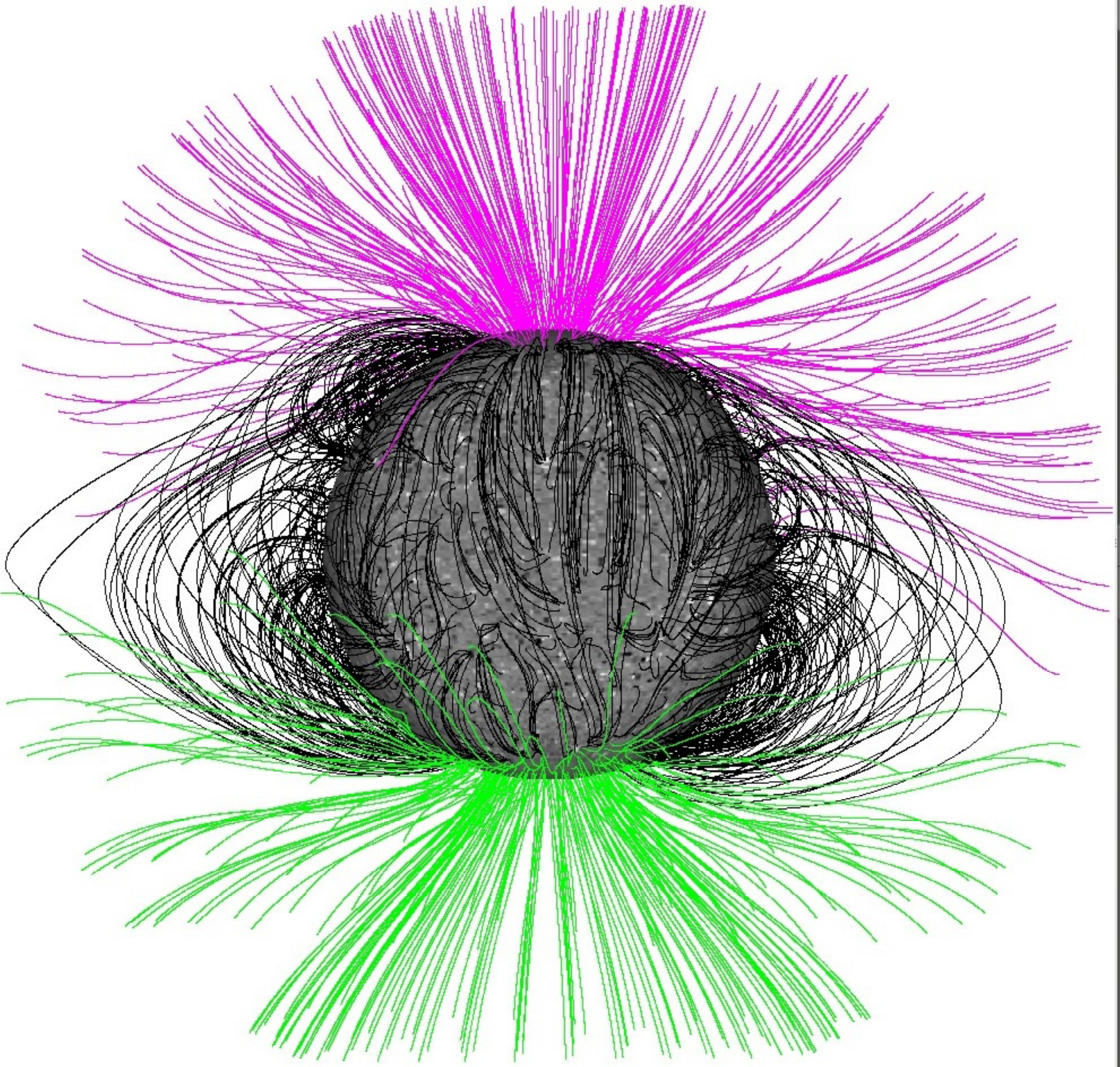}
\end{minipage}
\begin{minipage}{0.48\textwidth}
\includegraphics[width=\textwidth]{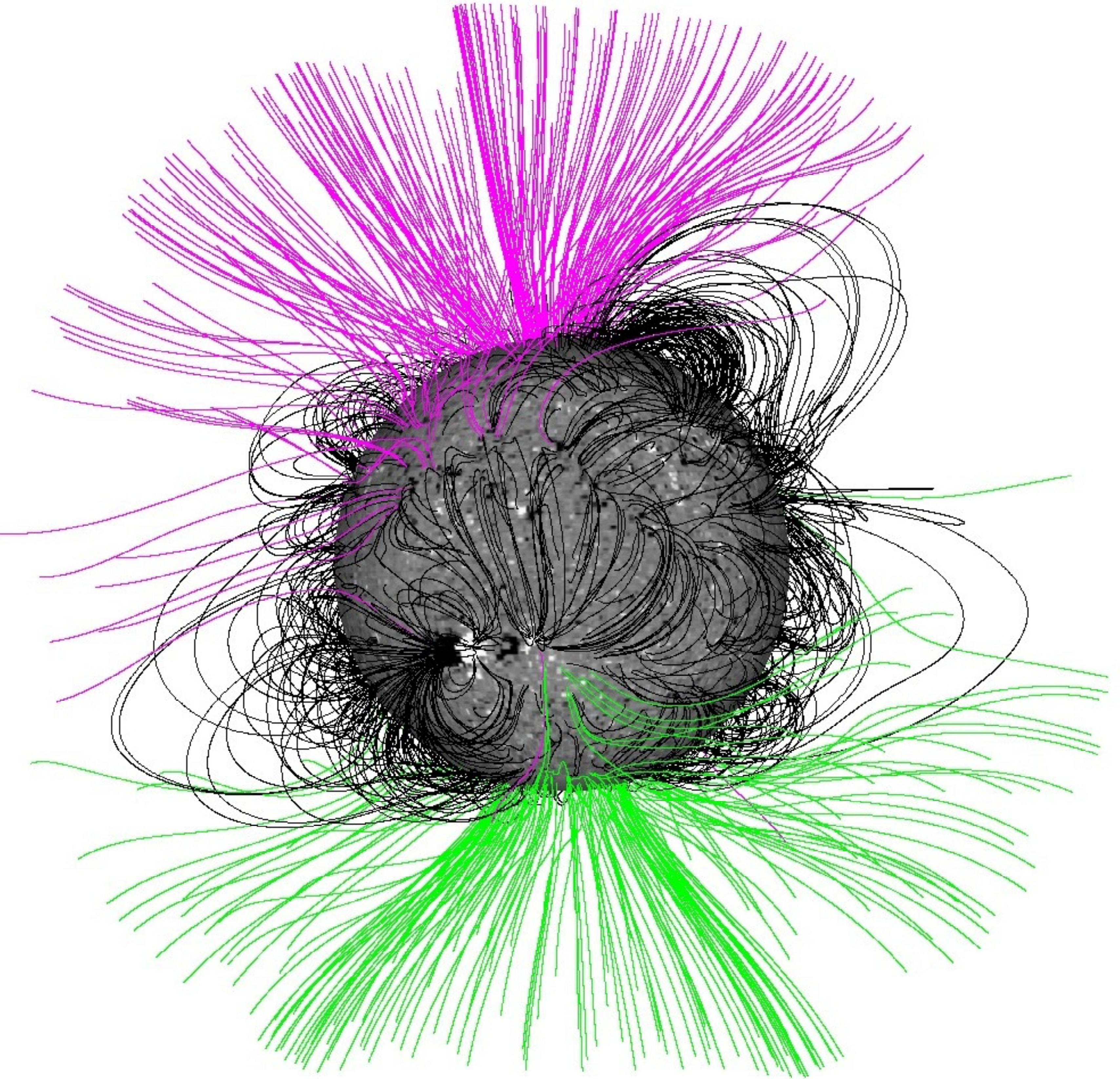}
\end{minipage}
\end{center}
\caption{{3D visualizations of the field lines of the PFSS models computed from MDI synoptic magnetograms for CR-2081 (left) and CR-2099 (right). Both models seen from 0º of latitude and 230º longitude.}\label{MagneticField}}
\end{figure*}

Figure \ref{MagneticField} shows 3D visualizations of the field lines of the PFSS models computed from MDI synoptic magnetograms for CR-2081 and CR-2099. For each traced field line the 3D coordinates of a starting point must be specified. In order to evenly cover the whole volume spanned by the DEMT reconstructions, one starting point was selected at the center of each tomographic cell at 10 uniformly spaced heights, ranging from 1.035 to 1.215\,R{$_\odot$}, and every 2$^\circ$ in both latitude and longitude, for a total of $162,000$ starting points and traced field lines. The geometry of the magnetic field line of the PFSS model passing through each starting point is then computed, both outward and inward, until {its footpoint (1.0\,R{$_\odot$})} and/or the source surface (2.5\,R{$_\odot$}) is reached.  To do so, the first order differential equations ${\rm d}r/B_r = r{\rm d}\theta/B_\theta = r\,\sin(\theta)\,{\rm d}\phi/B_\phi$ are numerically integrated using the PFSS Solarsoft package.

The next step is to trace the DEMT results along the computed magnetic field lines. To that  end, once the field line geometry in high resolution is completed, only one sample point per tomographic cell is kept, the median one. {To each sample point, the DEMT products ($N_e$, $T_m$, and $E_r$) in the tomographic cell where it is located are assigned to it, namely the electron density $N_e$, the electron mean temperature $T_m$, and the radiative power $E_r$.}

\begin{figure*} 
\begin{center}
\begin{minipage}{0.48\textwidth}
\includegraphics[width=\textwidth]{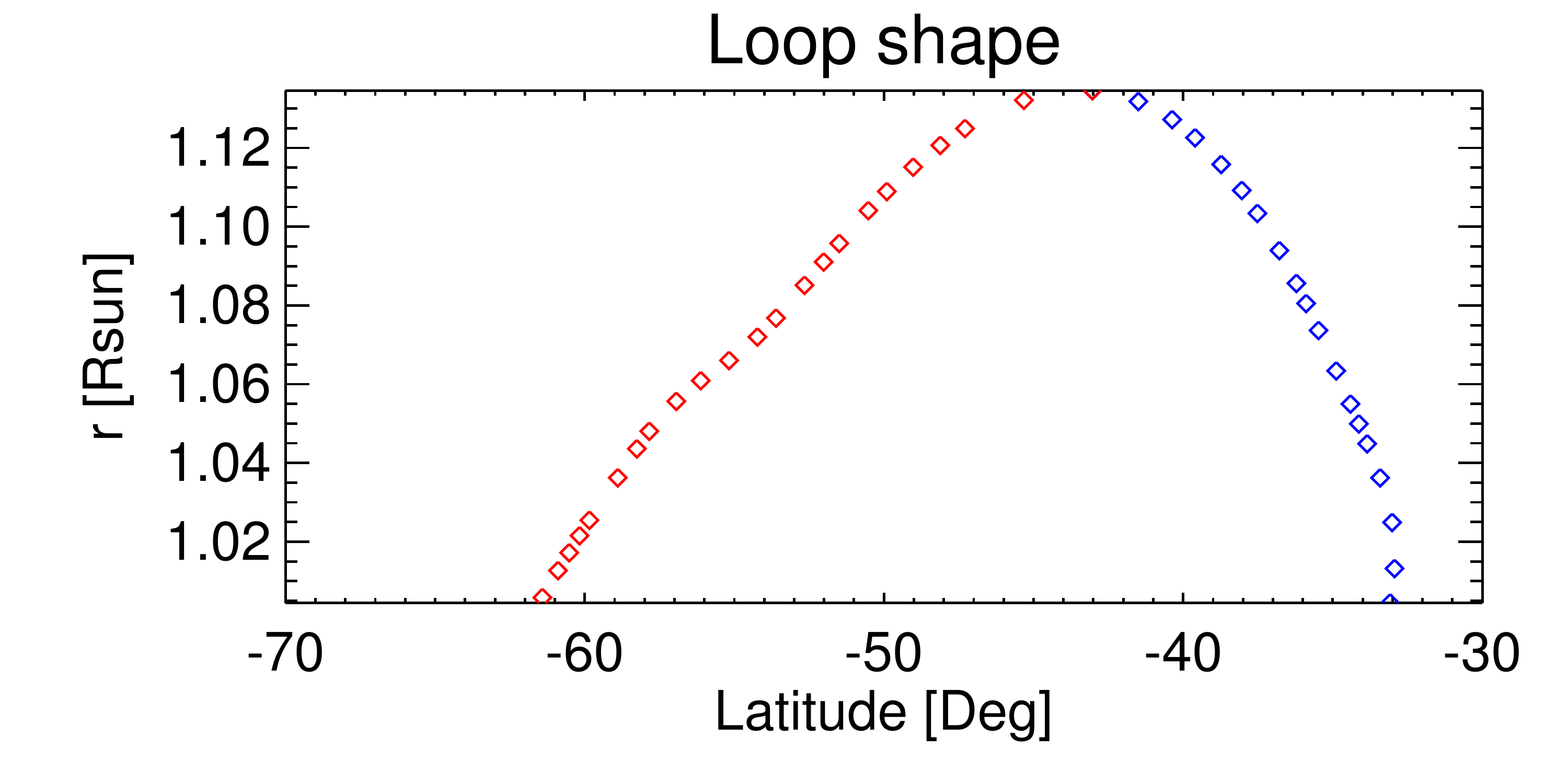}
\includegraphics[width=\textwidth]{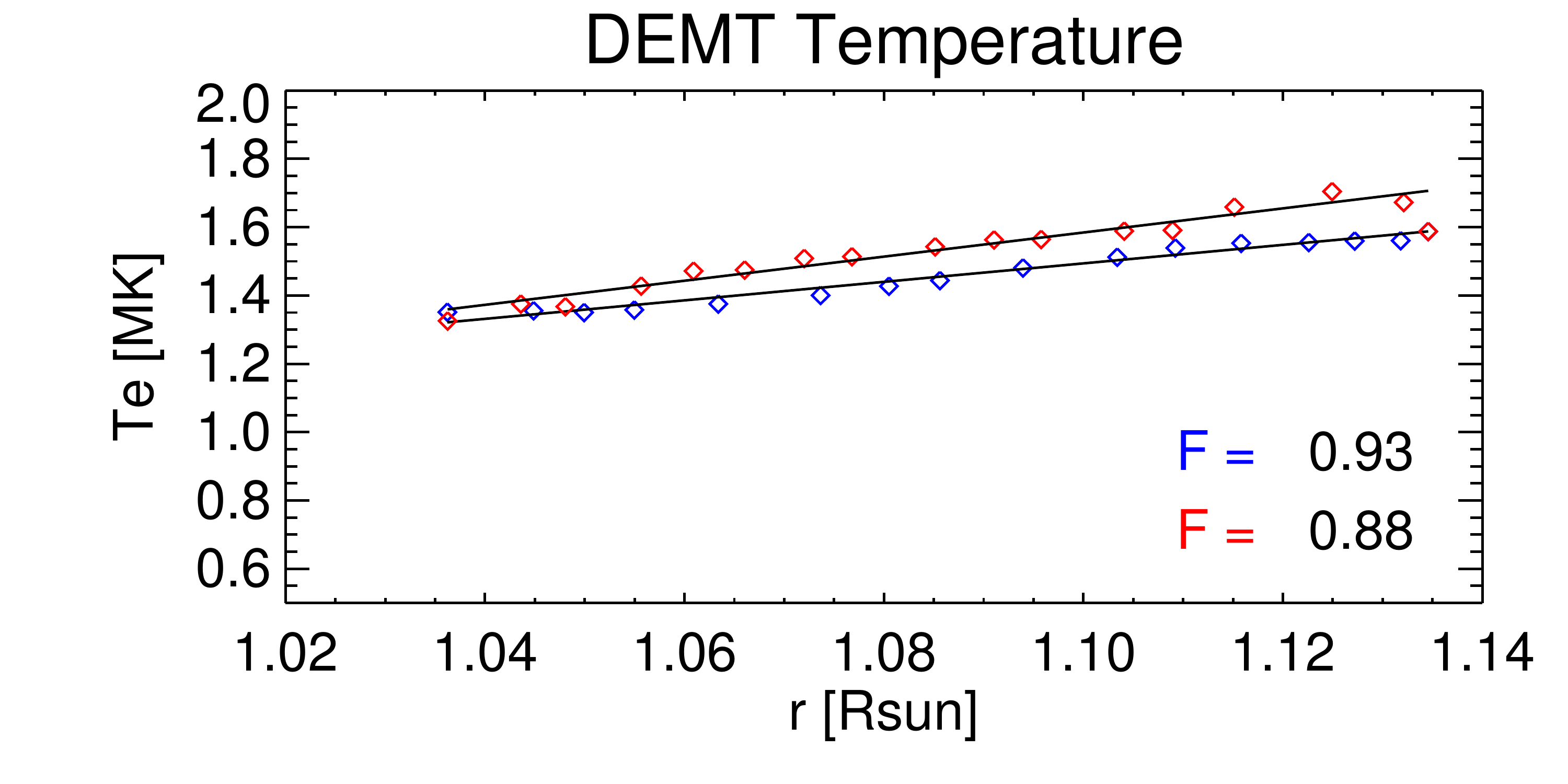}
\end{minipage}
\begin{minipage}{0.48\textwidth}
\includegraphics[width=\textwidth]{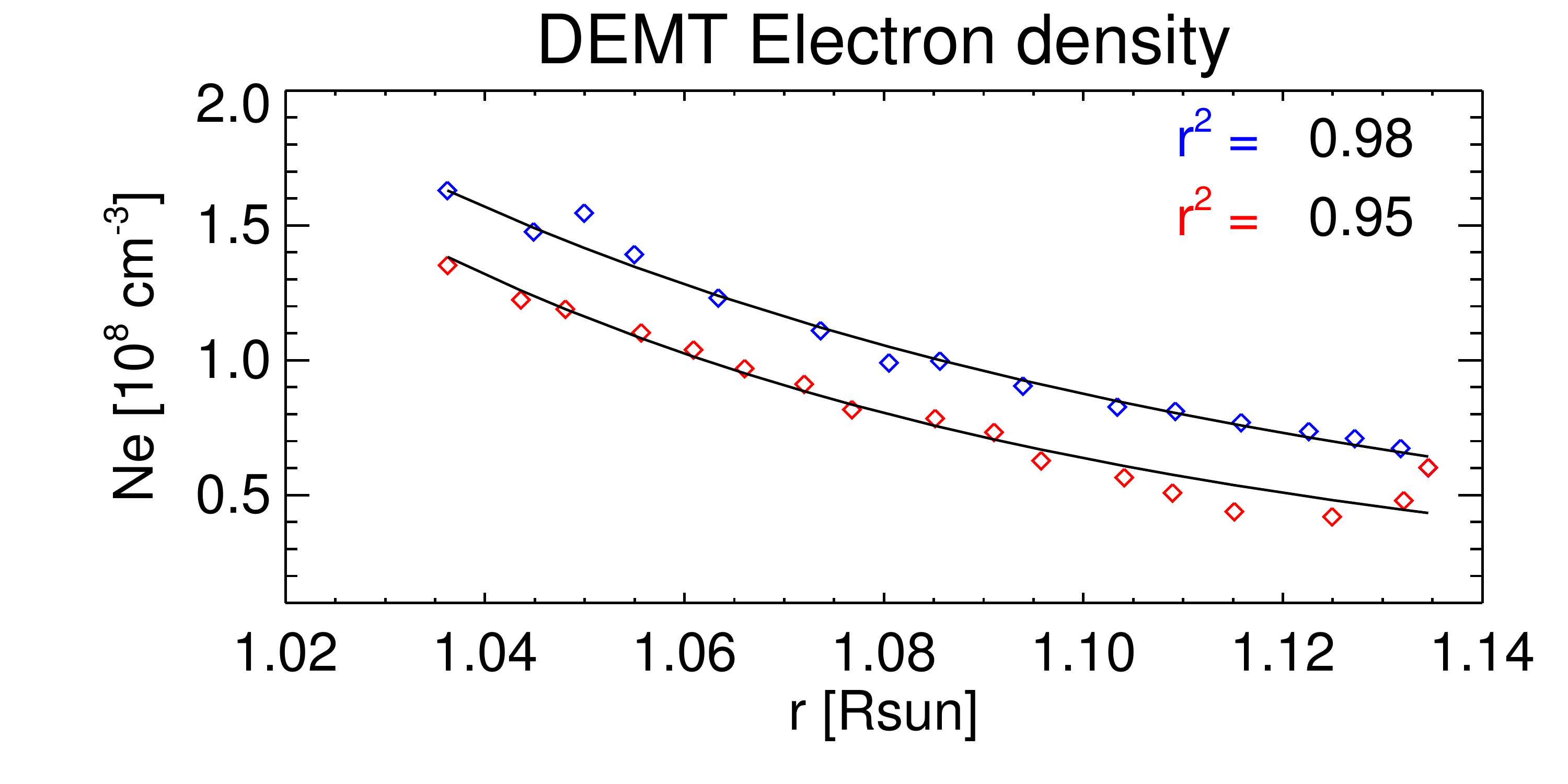}
\includegraphics[width=\textwidth]{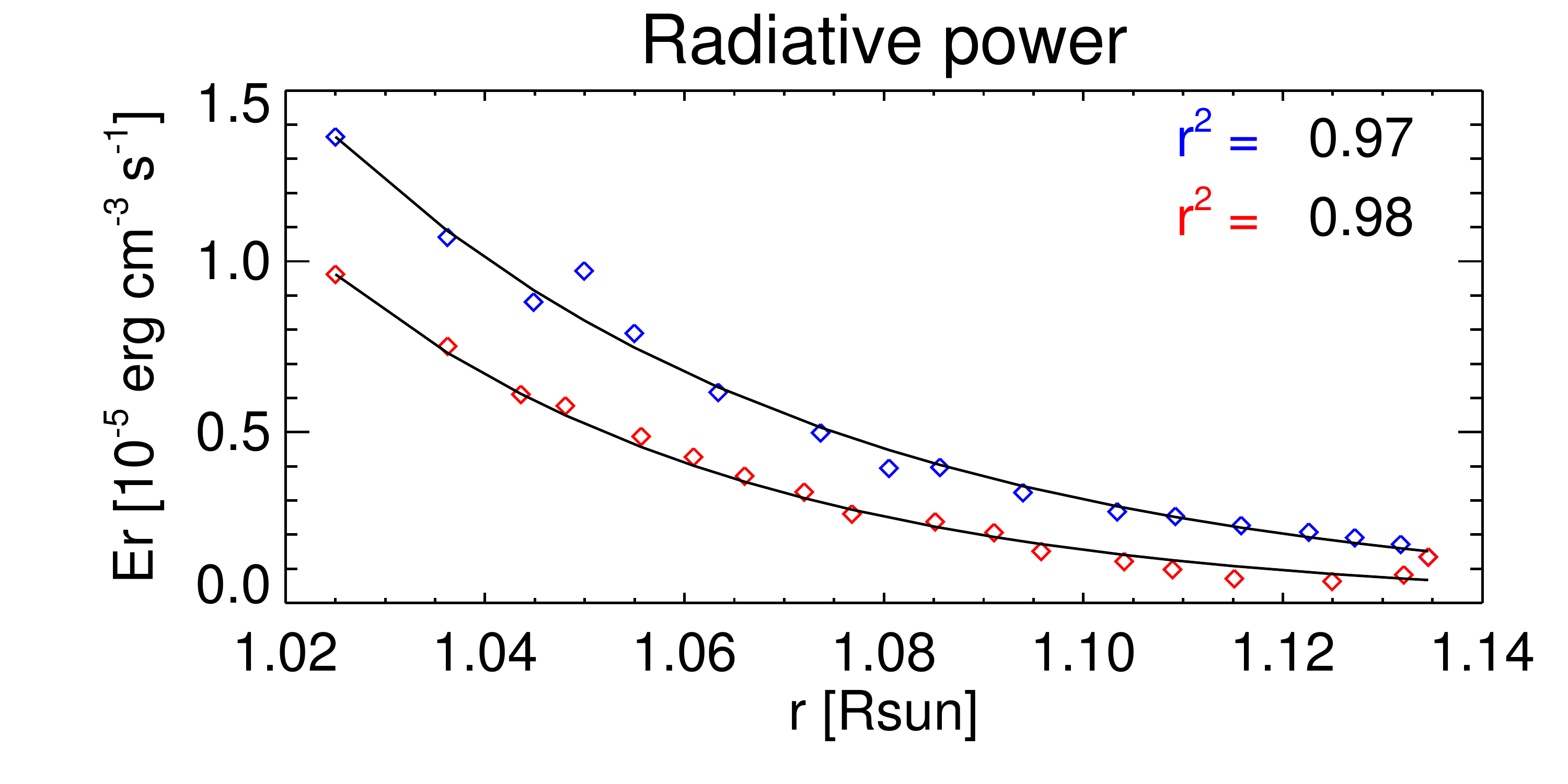}
\end{minipage}
\end{center}
\caption{DEMT results traced along magnetic loops of the PFSS model. The two legs of the loop are colored red and blue, with diamonds indicating data points. Top-left: loop shape projected in the radial-latitudinal plane. Top-right: electron density and exponential fits (see text). Bottom-left: electron temperature and linear fits (see text). Bottom-right: radiative power and exponential fits (see text). T he parameters $r^2$ and $F$ are explained in the text of Section \ref{criteria}. (A color version of this figure is available in the online journal).}
\label{trazados}
\end{figure*}

Figure \ref{trazados} shows an example along a closed magnetic loop. Note how the results are separated into the two \emph{legs} of the loop, defined as the two segments that go from the coronal base up to the apex of the loop. For each leg separately, exponential least-square fits are applied to both the density and the radiative power data points as a function of height. {Given their much less strong variation with height, temperature data points are fitted to a linear function. In this case the Theil-Sen estimator was preferred over} the least-square fit, being more robust to outliers. These are common in solar rotational tomography results, mainly due to the effect of unresolved coronal dynamics on the assumed static solution of the posed global optimization problem.

At this point, for every traced magnetic field line, its 3D geometry has been determined, the magnetic field strength along it $B(s)$ has been computed, and the radiative power $E_r(s)$ and electron mean temperature $T_m(s)$ have been determined from the fits to the traced DEMT data. As a result, all quantities involved in Equations (\ref{phir2}) and (\ref{phic2}) can be now numerically computed. Note that, as the quantity $\phi_c$ is sensitive to both the basal temperature and temperature gradient of the DEMT results (Equations (\ref{phic2}) and (\ref{Fc})), its computation from the fits to the traced DEMT data mitigates the effect of its stochasticity, mainly due to unresolved coronal dynamics. Once these quantities are known, the energy {input} flux $\phi_h$ is computed from Equation (\ref{FluxBalance}).

\subsection{Selection of Loops}\label{criteria}

The analysis of results is based on a selection of closed loops for which there are enough DEMT data points, they are evenly distributed over the range of heights spanned by the loop, and they are fairly described by their respective functional fits, as in the case shown in Figure \ref{trazados}.

As the electron density data points exhibits strong variations with height, the quality of their exponential fit is reasonably measured by its coefficient of determination $r^2$.\footnote{$r^2\equiv1-S_{\rm res}/S_{\rm tot}$, where $S_{\rm res}$ is the sum of the squared residuals and $S_{\rm tot}$ is the sum of data deviations from the mean.} In the case of the linear fit to the temperature, variations with height are sensibly smaller. Some loops may even be quasi-isothermal, and the coefficient of determination can be nearly zero, even for excellent fits, when the temperature gradient is low. Measuring the quality of the linear fit to the temperature based on the coefficient of determination, as done in previous works \citep{huang_2012,nuevo_2013}, would only select strong enough gradients. Interested in keeping loops with both strong and weak temperature gradients, {for the present study {these} criteria have been modified as described below}. 

A recent work by \citet{lloveras_2017} quantifies the impact of the main sources of systematic uncertainty of the DEMT technique into its products. In particular, the characteristic value of the temperature uncertainty is of order $\sim 5-10\%$ (depending on the coronal region). In order for a loop to be selected for analysis, the linear fit to the temperature is required to match the data within that uncertainty for a majority of the data points.

Based on the previous considerations, the numerical selection criteria listed below are based on actual experimentation with the data. These criteria aim at maximizing the {sample size}, while keeping only those loops for which the tomographic data can be fairly described by the exponential and linear fits to the electron density and temperature, respectively. As shown below, the selected sample size is larger than in previous studies, and evenly sample the coronal volume covered by the tomographic technique, resulting in a good representation of the complete tomographic results. Specifically, to be selected for analysis a closed loop must meet all following conditions:

\begin{enumerate}

\item 
Each leg of the loop must go through at least five tomographic grid cells with usable data (i.e. not labeled as ZDA or AEV), and there must be at least one data point in each third of the range of heights spanned by the loop.
 
\item 
The quality of the exponential fit to the density is $r^2>0.75$ in each leg of the loop.
 
\item
The linear fit to the temperature matches the DEMT values within their estimated error for at least a fraction $F>0.75$ of the data points in each leg of the loop.
 
\end{enumerate}

In Sections \ref{flux2081} through \ref{updown} below, the analysis is performed over all loops that meet the listed criteria.

\section{{Energy Flux Results}} \label{phis_results}  

{After analyzing all of the $162,000$ traced field lines in each rotation, about 54\% and 60\% are closed in CR-2081 and CR-2099, respectively. Some closed field lines do not have enough DEMT data points and/or are not well distributed, as specified by the first selection requirement listed above (Section \ref{criteria}), and some belong to ARs (as discussed in Section \ref{demt_results}). Those loops amount to about 17\% of the closed field lines in CR-2081, and to about 53\% for CR-2099 based on EUVI data, or 49\% based on AIA data.  On this remaining population the 2nd and 3rd selection criteria listed at the end of Section 4.3 are met by about 40\% of loops for CR-2081, 46\% of loops for CR-2099 based on EUVI, and 50\% of loops for CR-2099 based on AIA. The resulting number of loops is $\sim 29,000$ for CR-2081, and $\sim 21,000$ loops for CR-2099 based on EUVI, or $\sim 25,000$ based on AIA.}

Visual inspection of the temperature map of CR-2081 in Figure \ref{resultsDEMT} reveals that in the closed corona the low-latitudes are characterized by relatively cooler temperatures, while mid-latitudes are hotter. This is characteristic of the last solar minimum \citep{vasquez_2010,nuevo_2015}, as well as the previous period of minimum activity between SCs 22 and 23 \citep{lloveras_2017}. A similar behavior can be verified in the quiescent closed corona of CR-2099. In the analysis that follows these diverse thermodynamical regions are separated. To that end, magnetic loops were discriminated into those with both footpoints within a specific low-latitude range defined by $|{\rm latitude}| < 30^\circ$, and those within a mid-latitude range defined by $|{\rm latitude}| >30^\circ$. {After applying this selection criteria, the remaining population is $\sim 16,000$ for CR-2081, and $\sim 18,000$ or $\sim 23,000$ for CR-2099 when based on EUVI or AIA data, respectively.}

\subsection{Results for CR-2081} \label{flux2081}

\begin{figure*}%
\begin{center}
\begin{minipage}{0.48\textwidth}
\includegraphics[width=\textwidth]{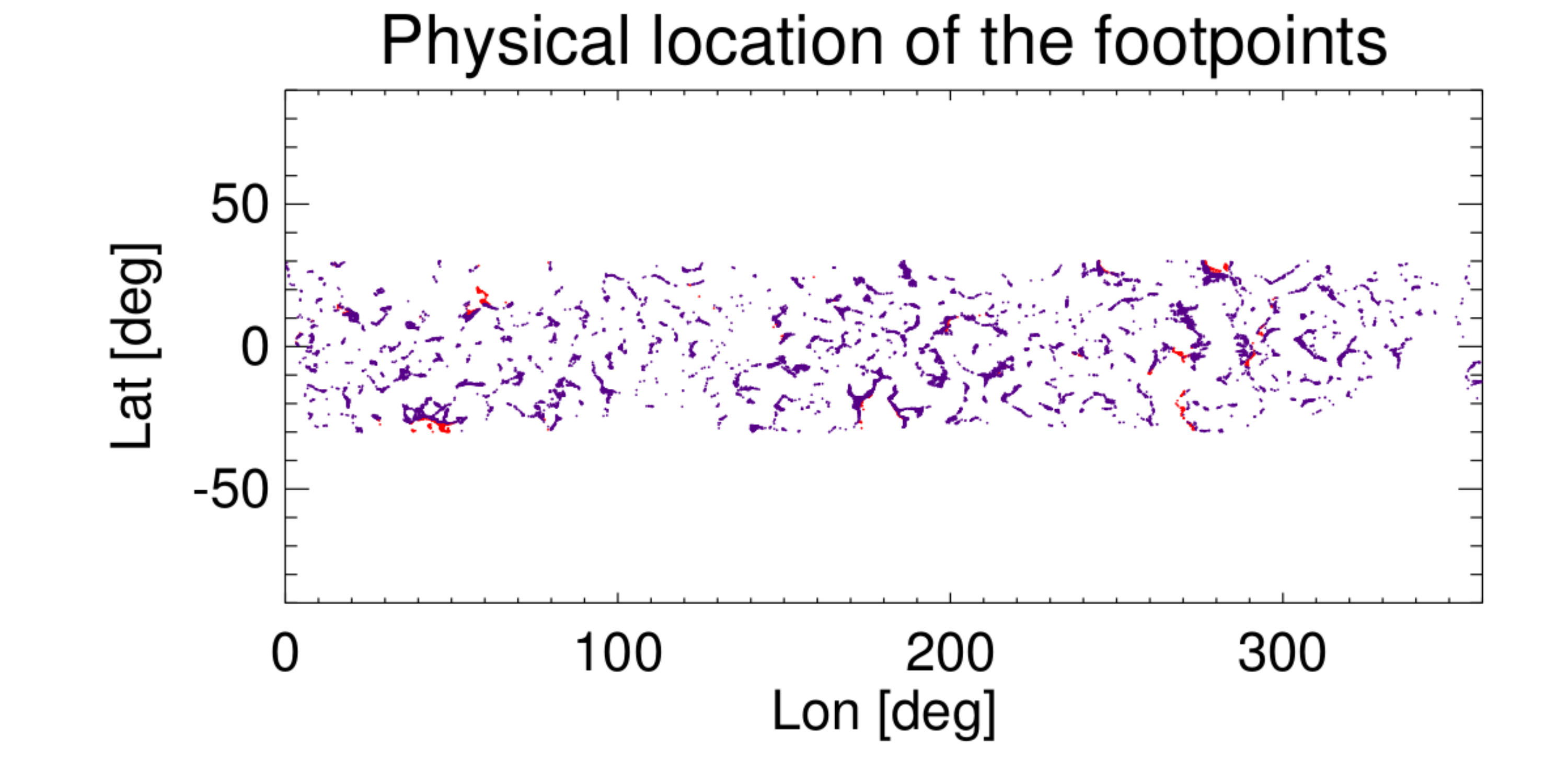}
\includegraphics[width=\textwidth]{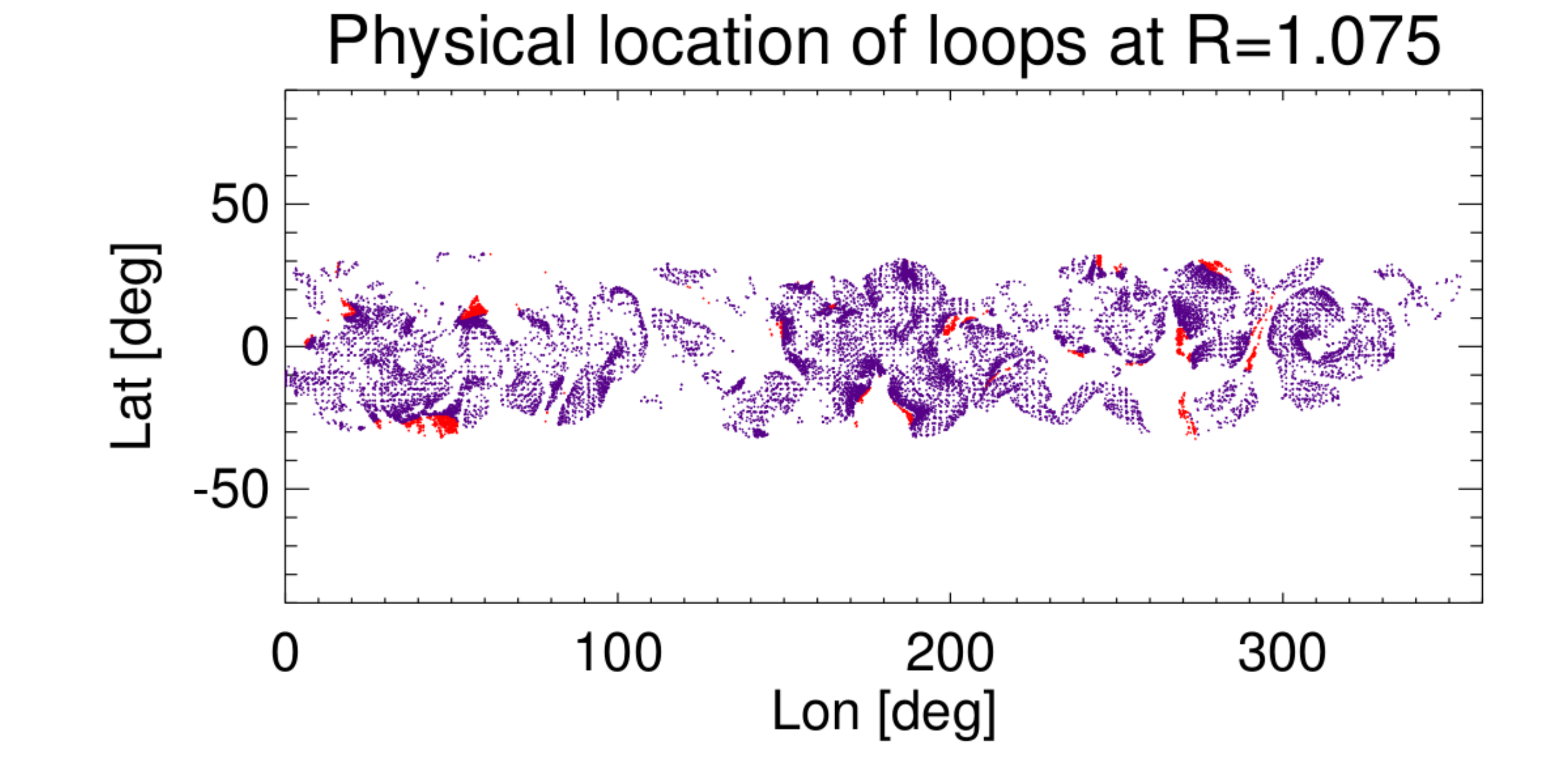}
\includegraphics[width=\textwidth]{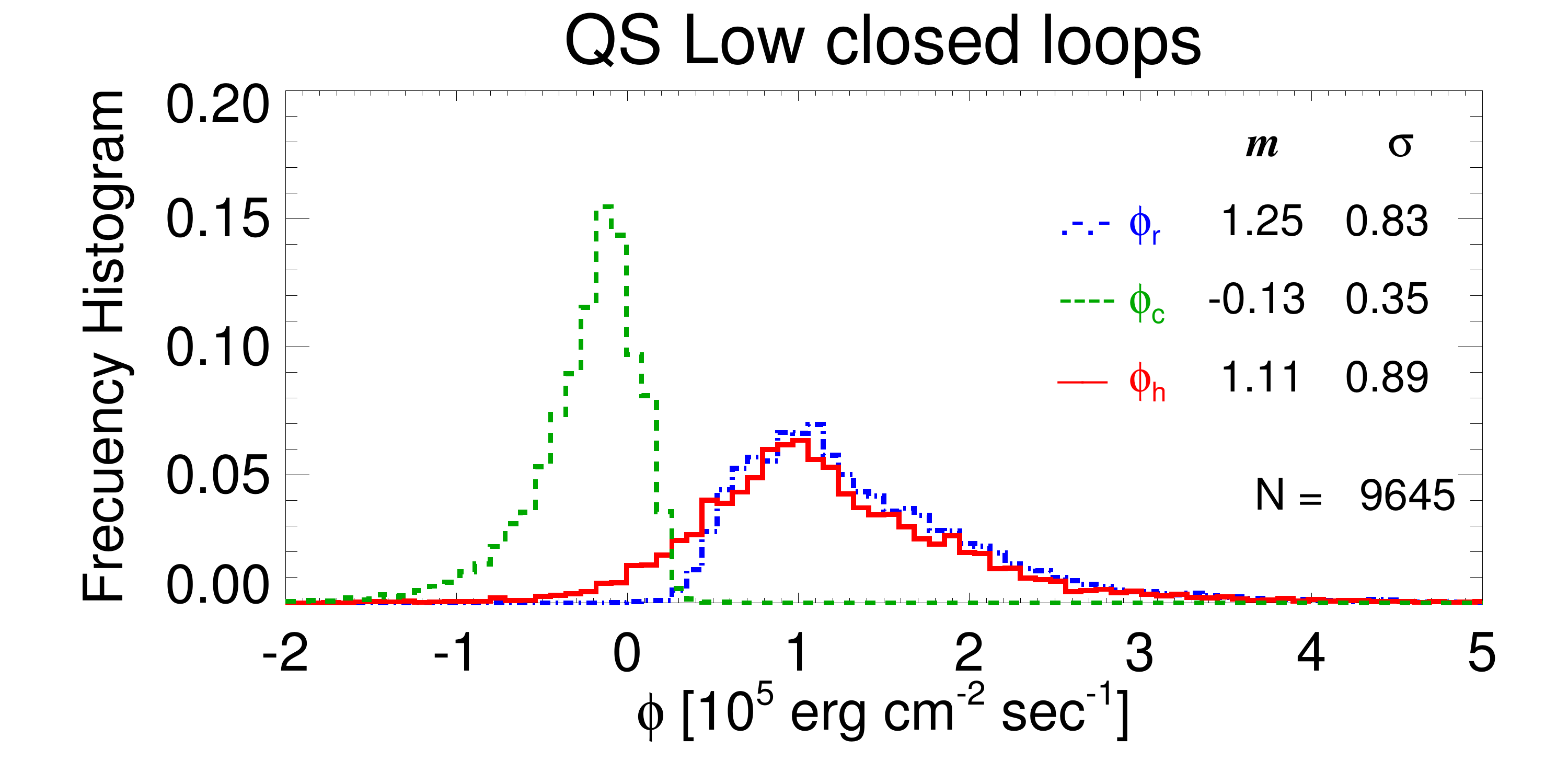}
\end{minipage}
\begin{minipage}{0.48\textwidth}
\includegraphics[width=\textwidth]{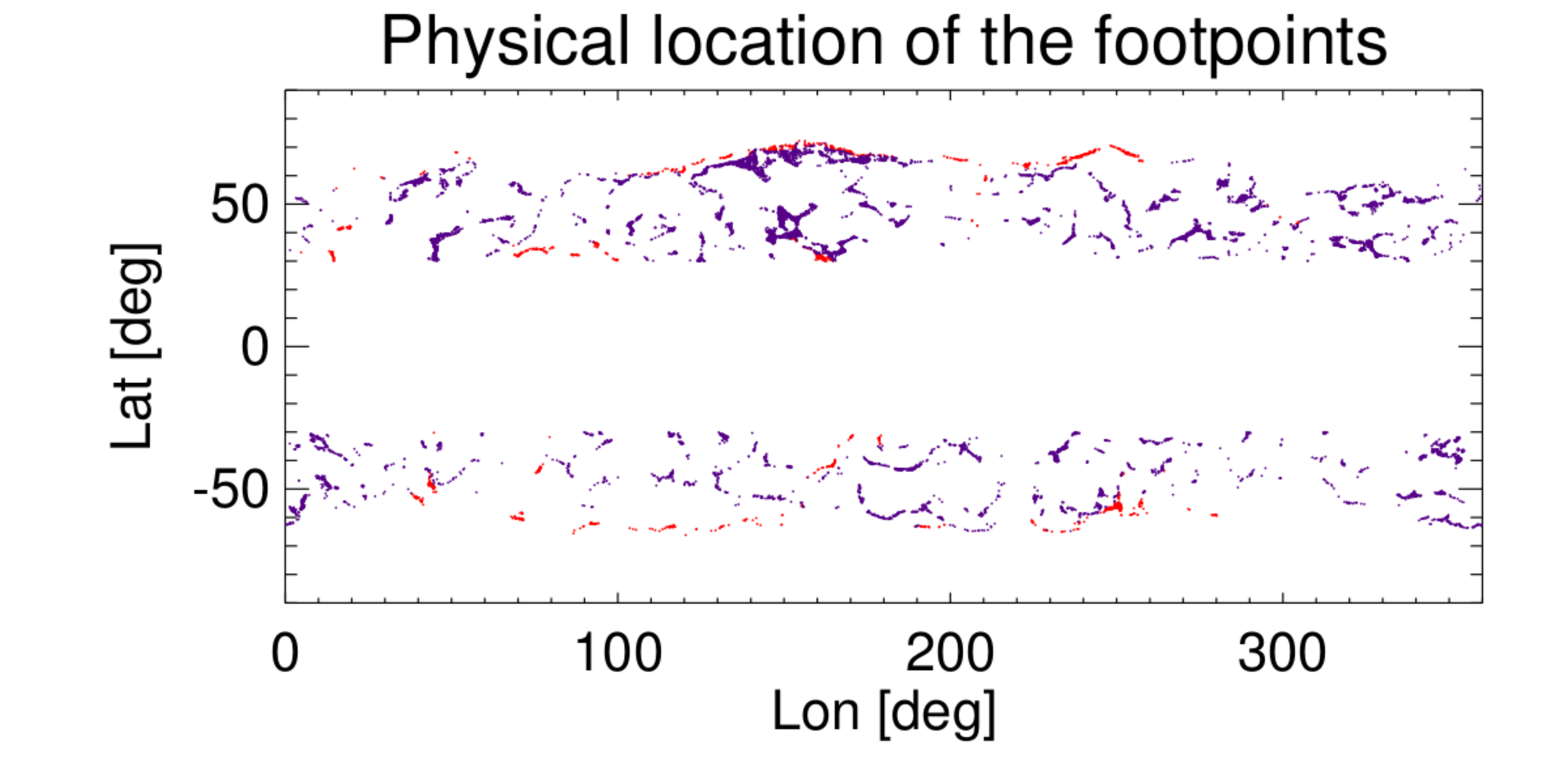}
\includegraphics[width=\textwidth]{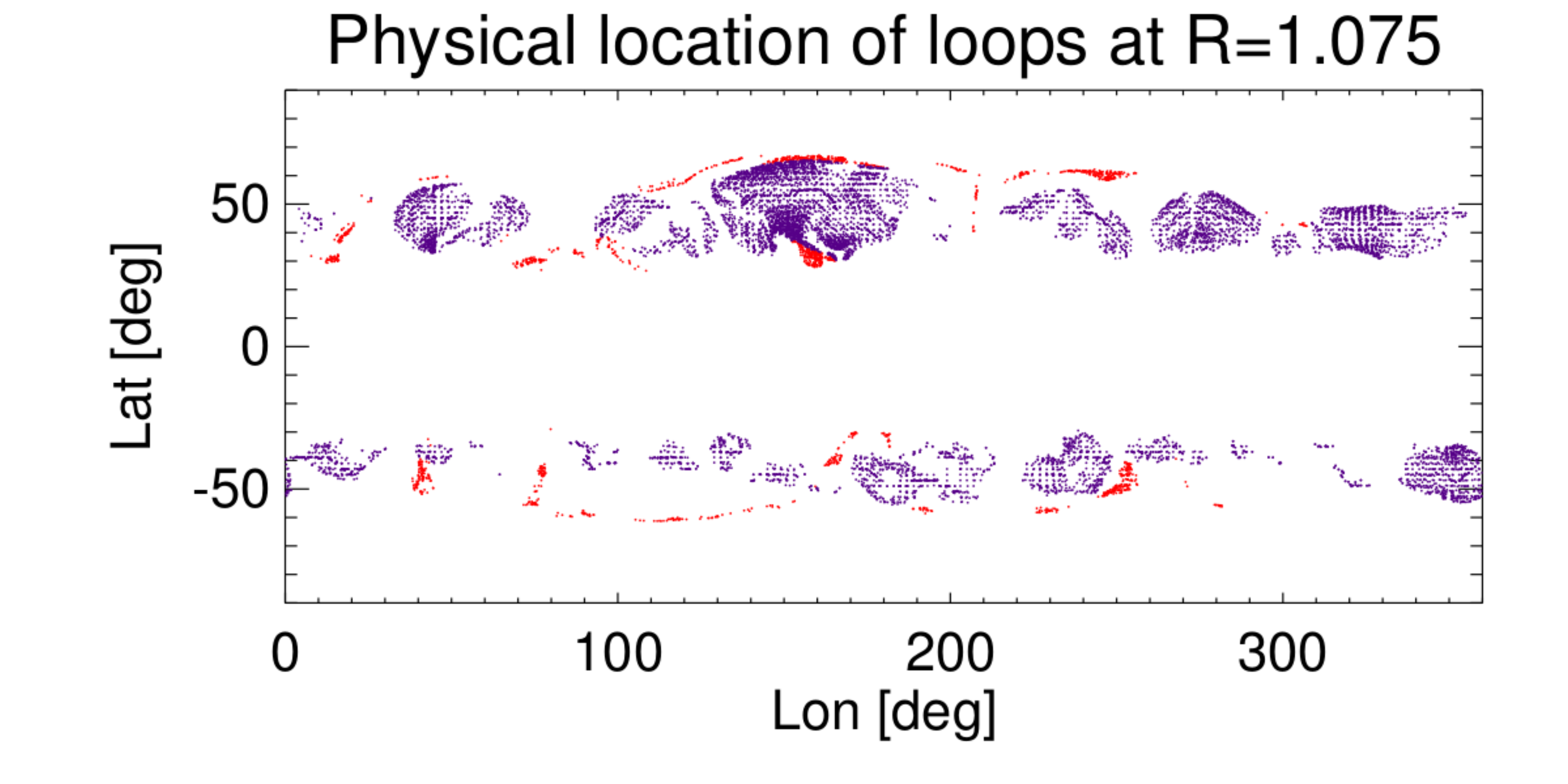}
\includegraphics[width=\textwidth]{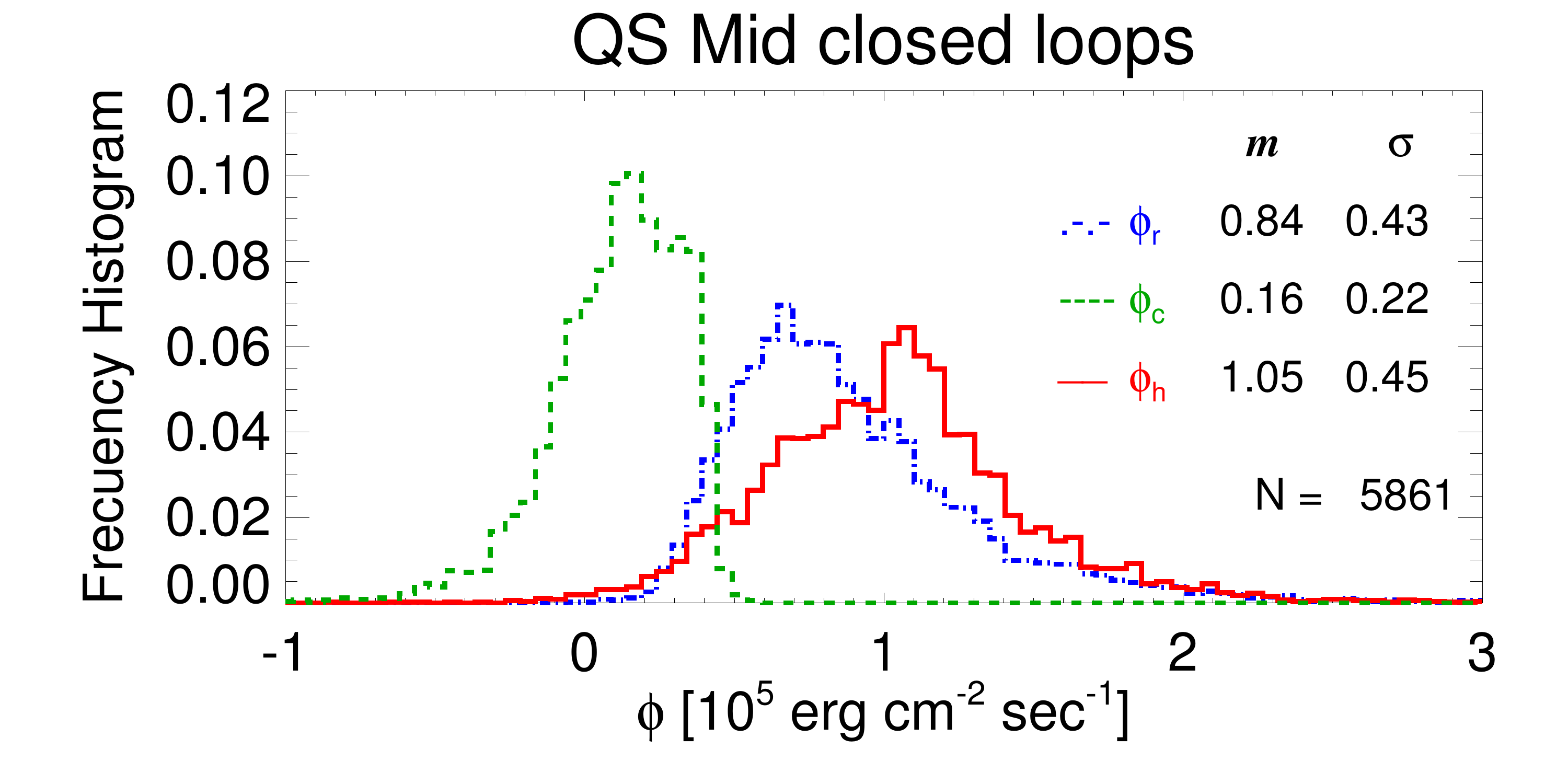}
\end{minipage}
\end{center}
\caption{Statistical results of energy flux quantities for CR-2081. Left/right panels correspond to the low/mid-latitude regions. Top: physical location of {the footpoints of the loops ($r=1.0\,{\rm R}_\odot$)}. Middle: physical location of the loops at a larger height ($r=1.075\,{\rm R}_\odot$). In the top and middle panels, the violet/red colored dots correspond to small/large loops (see text). Bottom: distribution of values of the loop-integrated quantities $\phi_r$, $\phi_c$, and $\phi_h$. The median $m$ and standard deviation $\sigma$ values of each distribution is tabulated, and the total number $N$ of analyzed loops is indicated. (A color version of this figure is available in the online journal).}
\label{flujo2081EUVI}
\end{figure*}

Results for CR-2081 are shown in Figure \ref{flujo2081EUVI}, where the left and right panels correspond to the low- and mid-latitude regions, respectively. The top panels show the spatial location of the footpoints of the loops {(i.e. at $r=1.0\,{\rm R}_\odot$)}, while the middle ones show their location at an intermediate height of the tomographic computational grid (at $r=1.075\,{\rm R}_\odot$). Comparison between top and middle panels gives a feeling of the divergence of the magnetic field lines. Loops for which the apex is within the range of heights covered by DEMT (dubbed as ``small'' loops hereafter) are indicated in violet color in the on-line version of the Figure, while those with a higher apex are indicated in red color (dubbed as ``large'' loops hereafter). For the selected loops, the bottom panels show the corresponding distribution of values of the loop-integrated quantities $\phi_r$, $\phi_c$, and $\phi_h$. The total number $N$ of analyzed loops is shown, along with the median $m$ and standard deviation $\sigma$ values of each distribution.

It is readily seen that the integrated radiative loss of the loops, measured by the quantity $\phi_r$, is larger in the low-latitudes. This is mainly due to the fact that, in the range of sensitivity of the EUVI instrument, namely 0.5-3.0 MK \citep[see]{nuevo_2015}, the radiative loss function $\Lambda(T)$ used in this work has a local maximum at $T \approx 1\,{\rm MK}$. The average temperature of the low- and mid-latitudes is {$1.17$ and $1.38$} MK, respectively (see Table \ref{comprot}), which explains a larger radiative loss at low-latitudes. This happens despite the fact that the average loop-length for the low- and mid-latitudes regions are $0.55$ and $0.74$ R$_\odot$, respectively, which implies a {larger length-integral} in the mid latitude loops. Still, most of the coronal radiative loss occurs at lower heights as $E_r\propto N_e^2$, which decays very rapidly with height.

\begin{table*}%
\begin{center}
\begin{tabular}{|c|c|c|c|c|c|c|c|}
\hline
  Instrument & Latitude & $N_{\rm tot}$ &$\left<\bar{N}_e\right> (\sigma)$ & $\left<\bar{T}_m\right> (\sigma)$ & $\left<\phi_r\right> (\sigma)$ & $\left<\phi_c\right> (\sigma)$ & $\left<\phi_h\right> (\sigma)$ \\ 
  \hline
  & & & $[10^8\,\rm{cm}^{-3}]$ & $[{\rm MK}]$ & \multicolumn{3}{|c|}{$[10^5\,\rm{erg\,cm^{-2}\,sec^{-1}}]$} \\
  
\hline \hline
   EUVI & Low    & 9645 & 0.93 (0.18) & 1.17 (0.10) & 1.25 (0.83) & -0.13 (0.35) & 1.11 (0.89) \\ 
        & Middle & 5861 & 0.99 (0.17) & 1.38 (0.11) & 0.84 (0.43) &  0.16 (0.22) & 1.05 (0.45) \\
\hline
\end{tabular}
\caption{Global statistics for the CR-2081 results, discriminating low and mid latitudes. For both populations the table shows the sample size $N_{\rm tot}$, and the median value (indicated as $\left<\,\right>$) and standard deviation ($\sigma$) of the height-averaged DEMT electron density $\bar{N}_e$ and temperature $\bar{T}_m$, as well as of the energy flux quantities $\phi_r$, $\phi_c$, and $\phi_h$.}
\label{comprot}
\end{center}
\end{table*}

Statistical results for CR-2081 are shown in Table \ref{comprot}, discriminating low- and mid-latitude loops. For both populations, the table shows the median value (indicated as $\left<\,\right>$) and the standard deviation ($\sigma$) of the flux quantities, as well as of the characteristic DEMT electron density $\bar{N}_e$ and temperature $\bar{T}_m$ of the loops, where the bar indicates the height-averaged value for each loop. 

While the quantity $\phi_r$ is defined positive, the conductive flux quantity $\phi_c$ is not. Note that low-latitudes are dominated by $\phi_c<0$ values, while mid-latitudes by $\phi_c>0$. The sign of $\phi_c$ is closely related to the temperature gradient with height. From Equation (\ref{Fc}) it can be easily shown that in magnetic loops for which $\phi_c<0$ the temperature decreases with height, while the opposite holds when the temperature increases with height. 

{Coronal magnetic structures for which the temperature increases/decreases with height {in the range of heights covered by DEMT, 1.02 to 1.22 ${\rm R}_\odot$,} have been dubbed as ``up"/``down" loops by \citet{huang_2012} and \citet{nuevo_2013}, who first observed their presence by means of DEMT. As speculated by the authors of those works, loops of type down can be expected if the heating deposition is strongly confined near the coronal base of a magnetic loop. Down loops were first predicted by \citet{serio_1981}, and later on by \citet{aschwanden_2002}. {In a recent study by \citet{schiff_2016}, down and up loops have been successfully reproduced by a numerical implementation of a 1D steady state model that considers time-averaged heating rates}. The analysis of these structures in the context of the new tool here developed is shown in Section \ref{updown} below.}

The resulting distributions of the energy input flux at the coronal base $\phi_h$ have similar median values in both regions, with a smaller standard deviation in the mid latitudes. The characteristic range of values considering both regions is $\phi_h \sim 0.5-1.5 \,\times 10^5\,\rm{erg\,cm^{-2}\,sec^{-1}}$.

\subsection{Results for CR-2099} \label{flux2099}

\begin{figure*} 
\begin{center}
\begin{minipage}{0.48\textwidth}
\includegraphics[width=\textwidth]{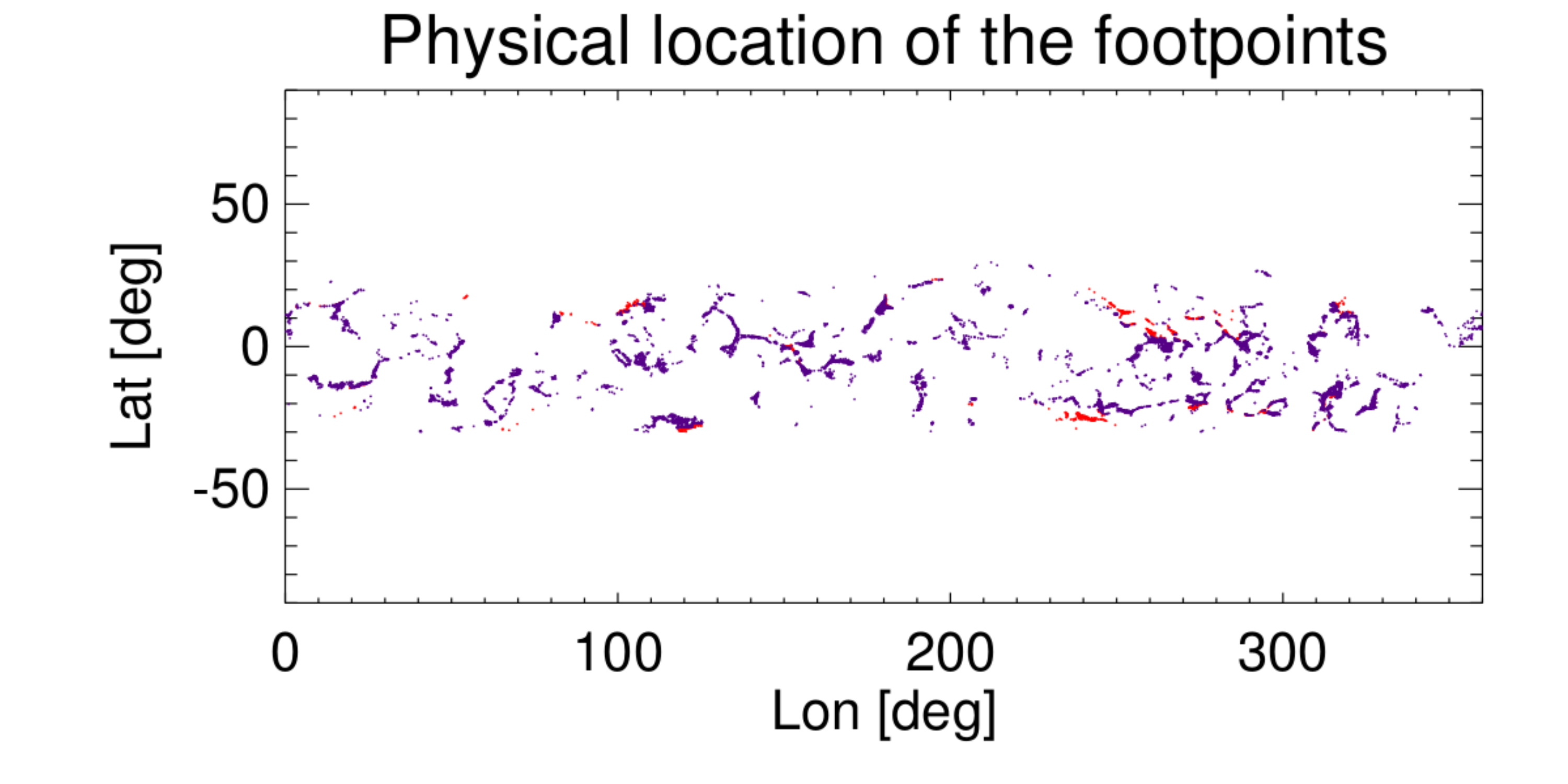}
\includegraphics[width=\textwidth]{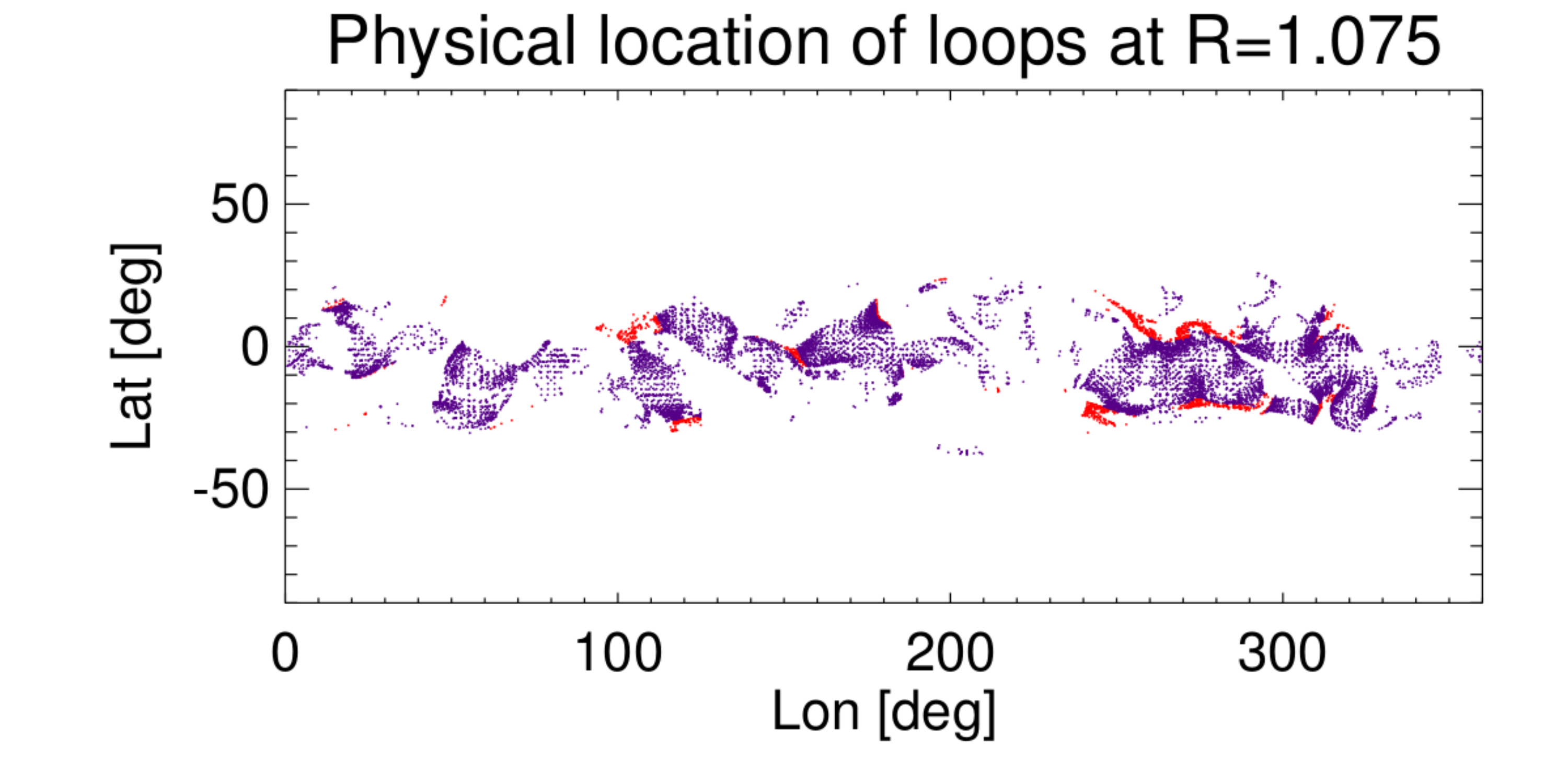}
\includegraphics[width=\textwidth]{f5e.pdf}
\end{minipage}
\begin{minipage}{0.48\textwidth}
\includegraphics[width=\textwidth]{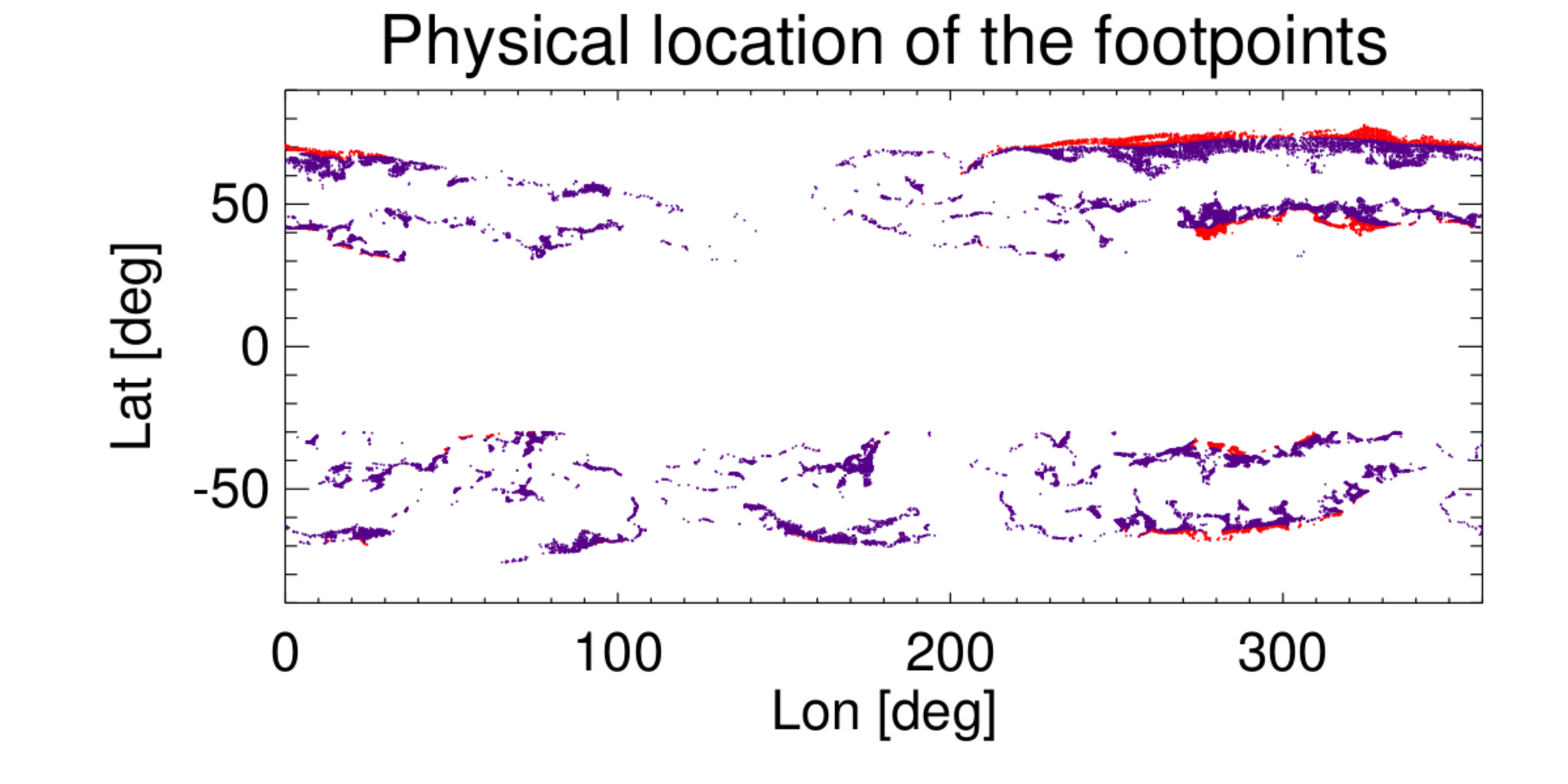}
\includegraphics[width=\textwidth]{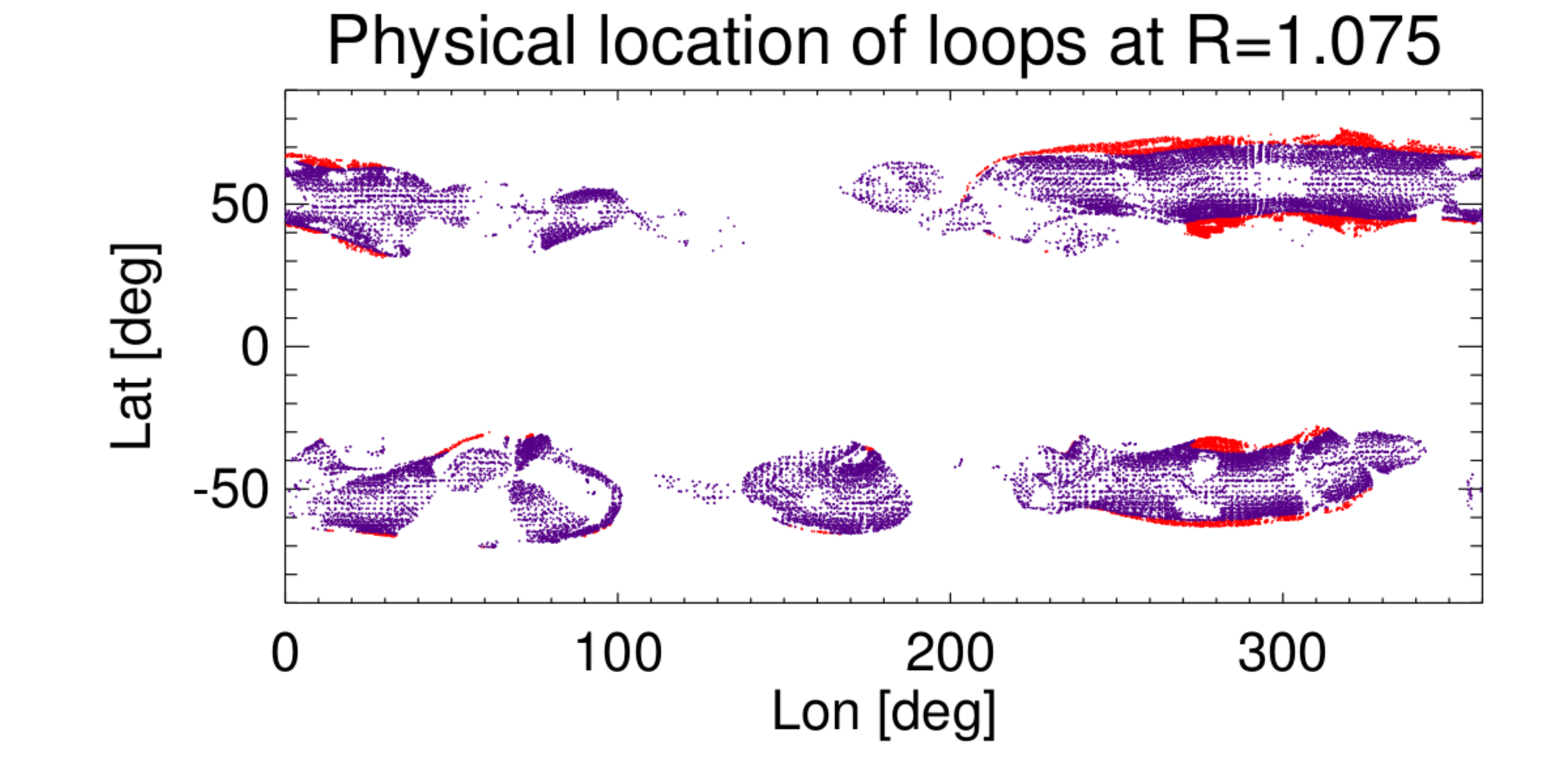}
\includegraphics[width=\textwidth]{f5f.pdf}
\end{minipage}
\end{center}
\caption{Similar to Figure \ref{flujo2081EUVI}, but for CR-2099 based on DEMT reconstructions from AIA data. (A color version of this figure is available in the online journal).}
\label{flujo2099AIA}
\end{figure*}

Figure \ref{flujo2099AIA} shows the results for CR-2099 based on AIA data, where the left and right panels correspond to low- and mid-latitude regions, respectively. The same analysis was also performed for this rotation based on EUVI data, leading to results consistent to those based in AIA data. Table \ref{CompEUVIAIA} details the statistical results for CR-2099 based on both data sets. The results obtained with both instruments are more consistent in mid-latitudes, due to the similar sample size. In the case of low latitudes, the EUVI sample size is considerably smaller (about half) than for AIA. Differences in sample size are due to the same selection criteria being applied to different data sources, but the precise reasons for which the AIA based analysis is able to retrieve a larger data sample is not clear. In any case, results from both data sets lead to similar characteristic distributions.

\begin{table*} 
\begin{center}
\begin{tabular}{|c|c|c|c|c|c|c|c|}
\hline
  Instrument & Latitude & $N_{\rm tot}$ &$\left<\bar{N}_e\right> (\sigma)$ & $\left<\bar{T}_m\right> (\sigma)$ & $\left<\phi_r\right> (\sigma)$ & $\left<\phi_c\right> (\sigma)$ & $\left<\phi_h\right> (\sigma)$ \\ 
  \hline
  & & & $[10^8\,\rm{cm}^{-3}]$ & $[{\rm MK}]$ & \multicolumn{3}{|c|}{$[10^5\,\rm{erg\,cm^{-2}\,sec^{-1}}]$} \\
  
\hline \hline
  EUVI & Low & 3243 & 0.89 (0.15) & 1.40 (0.16) & 1.19 (0.81) & 0.15 (0.31) & 1.35 (0.83) \\ 
   & Middle & 14724 & 0.83 (0.12) & 1.60 (0.10) & 0.83 (0.35) & 0.42 (0.20) & 1.31 (0.34) \\
\hline
  AIA & Low & 6891 & 0.92 (0.17) & 1.47 (0.11) & 1.08 (0.84) & 0.03 (0.48) & 1.12 (0.94) \\ 
  & Middle & 15820 & 0.86 (0.14) & 1.61 (0.11) & 0.77 (0.33) & 0.44 (0.28) & 1.27 (0.41) \\
\hline
\end{tabular}
\caption{Similar to Table \ref{comprot}, but for CR-2099 based on DEMT reconstructions from both EUVI and AIA data, alternatively.}
\label{CompEUVIAIA}
\end{center}
\end{table*}

Comparing both rotations, in the low-latitudes the results are similar, being the most notable difference that the distribution of values of the quantity $\phi_c$ for CR-2099 is not dominated by negative values as it is for CR-2081. This is consistent with the finding by \citet{nuevo_2013} that down loops are prominent during solar minimum. In the mid-latitudes of CR-2099, $\phi_c$ is virtually positive everywhere, which is consistent with the fact that down loops not only diminish in number with increasing activity but also tend to be found only at low-latitudes as activity increases \citep{nuevo_2013}. It is also to be noted an increase of the characteristic values of the input energy flux $\phi_h$ for CR-2099 compared to CR-2081, specially at mid-latitudes where it shows a $\sim 20\%$ larger median value. This is consistent with the relatively higher temperatures in the mid-latitude regions for CR-2099 (see bottom panels in Figure \ref{resultsDEMT}).

\subsection{Analysis of Temperature Structures} \label{updown}

\begin{figure*} 
\begin{center}
\begin{minipage}{0.48\textwidth}
\includegraphics[width=\textwidth]{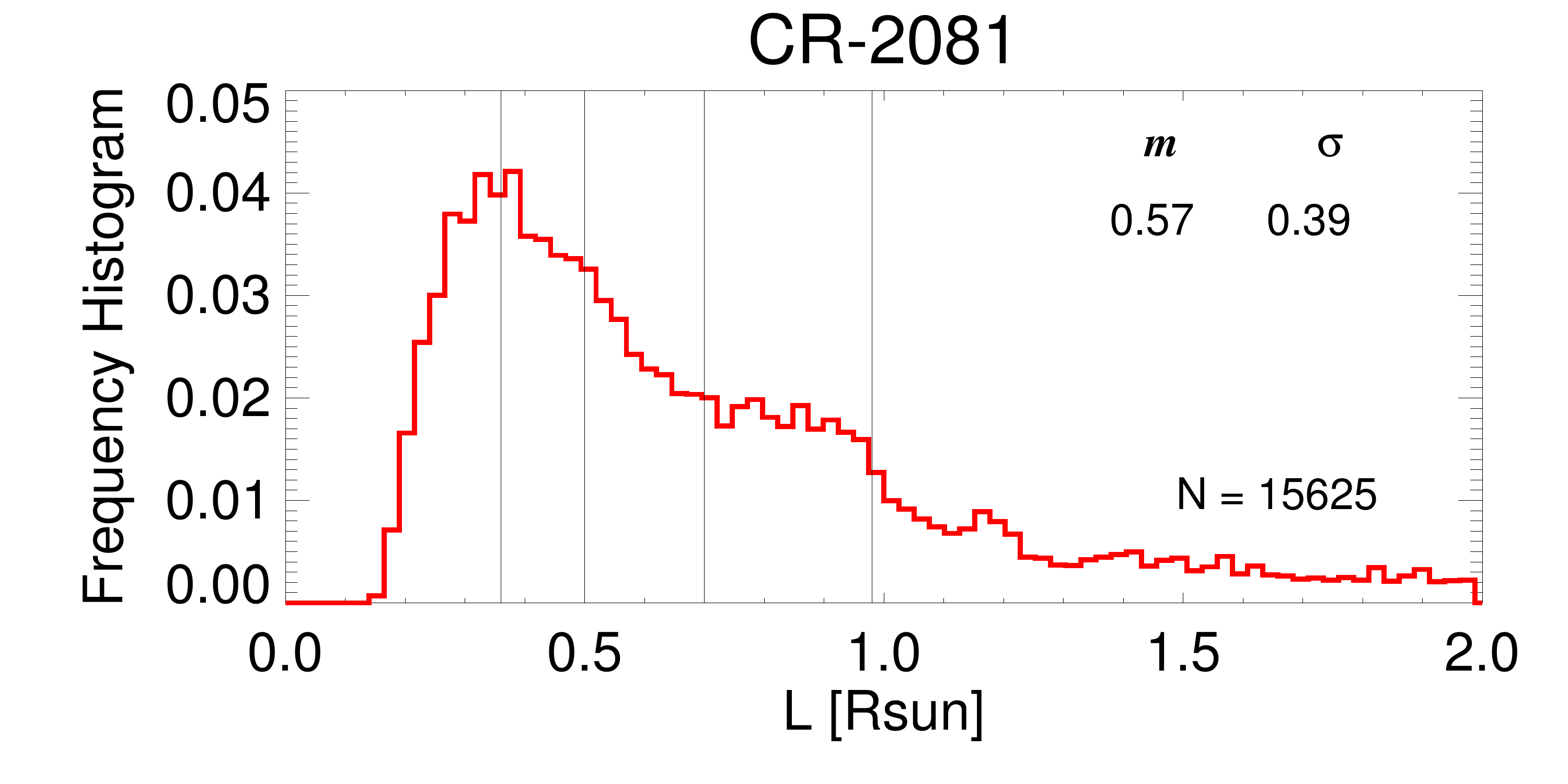}
\end{minipage}
\begin{minipage}{0.48\textwidth}
\includegraphics[width=\textwidth]{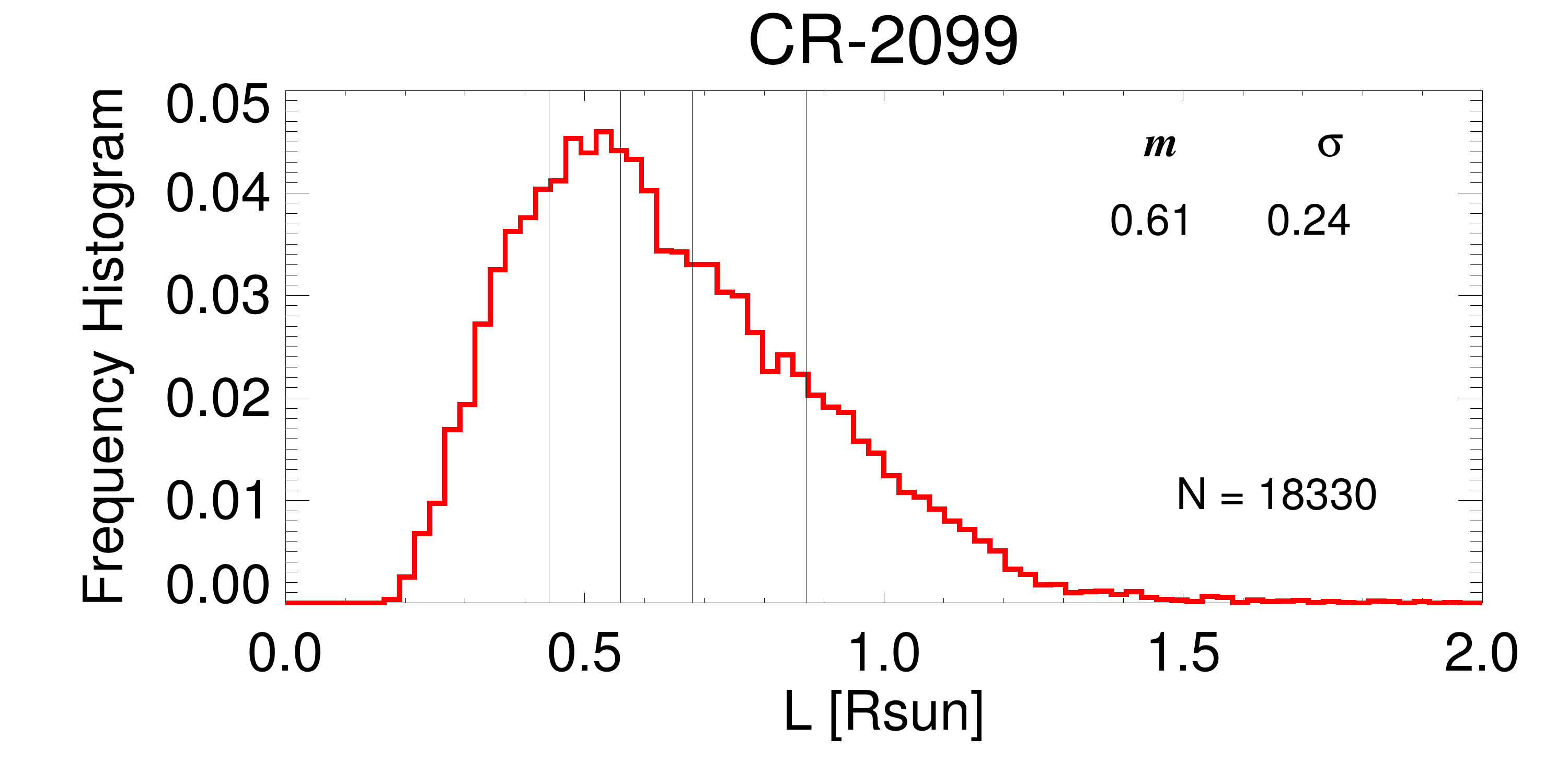}
\end{minipage}
\end{center}
\caption{Histograms of length $L$ of the up and down loops for CR-2081 (left panel) and CR-2099 (right panel). The vertical lines indicate bin limits, defined for the study as a function of loop length.\label{histololos}}
\end{figure*}

As discussed in Section \ref{flux2081}, the sign of the conductive flux quantity $\phi_c$ is related to that of the temperature gradient with height. Down-loops, i.e., those for which the temperature decreases with height, are characterized by $\phi_c<0$, while the opposite holds for up-loops. While in previous DEMT works \citep{huang_2012,nuevo_2013} the loop-selection criteria focused only on down/up loops, our criteria in Section \ref{criteria} allow selection not only of down and up loops, but also of quasi-isothermal structures. 

A loop can be regarded as quasi-isothermal in the coronal region studied by DEMT if the temperature gradient is weak enough compared to the range of coronal heights $\delta r$ spanned by the loop. More specifically, $\delta r = r_{\rm max} - r_{\rm base}$, where $r_{\rm max} = r_{\rm appex}$ for small loops, and  $r_{\rm max} = 1.25\,{R_{\odot}}$ (the maxi\-mum height of the tomographic computation ball) for large loops. In this study a loop is then classified as quasi-isothermal if $|dT/ds| < \Delta(T)\,/\,\delta r$, where $dT/ds$ is the temperature gradient of the linear fit to the temperature data points, and $\Delta(T)$ is the characteristic uncertainty in temperature data points due to systematic errors, which is in the range $5-10\%$ as shown in \citet{lloveras_2017}. Down/up loops can then be separated by requesting $|dT/ds| > \Delta(T)\,/\,\delta r$, and further discriminated in their two populations according to the sign of the gradient. 

This classification criterion was applied to all closed loops analyzed in Sections \ref{flux2081} and \ref{flux2099} to separate the up and down loops. The loops {were then} further classified according to their length $L$. For both analyzed rotations, Figure \ref{histololos} shows histograms of the loop length $L$ of all up and down loops. The vertical lines indicate the limits of five loop length bins set to have a similar sample size within each bin. Tables \ref{longEUVI} and \ref{longAIA} show the statistical results of the whole population as well as for each bin separately. 

\begin{table*}%
\begin{center}
\begin{tabular}{|c|c|c|c|c|c|c|c|}
\hline
  $L_{\rm min}$ & $L_{\rm max}$ & $\left< L \right>$ & $N_{\rm down}$ & $N_{\rm tot}$ & $\frac{N_{\rm down}}{N_{\rm tot}}$ & $\left<\lambda_N\right>$ & $\frac{\left< L \right>/2}{\left<\lambda_N\right>}$ \\     
\hline \hline
  0.10 & 2.00 & 0.46 & 6600 & 15625 & 0.42 & 0.082 & 2.82 \\ 
\hline
  0.10 & 0.36 & 0.28 & 2144 &  3202 & 0.67 & 0.081 & 1.75 \\
  0.36 & 0.50 & 0.42 & 1521 &  3185 & 0.48 & 0.080 & 2.63 \\
  0.50 & 0.70 & 0.58 & 1200 &  3141 & 0.38 & 0.079 & 3.67 \\
  0.70 & 0.98 & 0.85 &  802 &  3156 & 0.25 & 0.085 & 5.00 \\
  0.98 & 2.00 & 1.43 &  933 &  2941 & 0.32 & 0.089 & 8.01 \\
\hline
\end{tabular}
\caption{Statistics down and up loops of CR-2081, discriminating different ranges $[L_{\rm min},L_{\rm max}]$ of the loop length $L$. For each range of lengths, the median value of loop length $\left< L \right>$ {is tabulated}, the number of down loops $N_{\rm down}$, the total number of loops $N_{\rm tot}$ (up and down), the median value of density scale height $\left< \lambda_N \right>$ of loops, and two ratios of discussed in the text.}
\label{longEUVI}
\end{center}
\end{table*}
  
\begin{table*}%
\begin{center}
\begin{tabular}{|c|c|c|c|c|c|c|c|}
\hline 
  $L_{\rm min}$ & $L_{\rm max}$ & $\left< L \right>$ & $N_{\rm down}$ & $N_{\rm tot}$ & $\frac{N_{\rm down}}{N_{\rm tot}}$ & $\left<\lambda_N\right>$ & $\frac{\left<L\right>/2}{\left<\lambda_N\right>}$ \\     
  \hline \hline
  0.10 & 2.00 & 0.55 & 1657 & 18133 & 0.09 & 0.081 & 3.40 \\
\hline
  0.10 & 0.44 & 0.35 &  494 &  3658 & 0.14 & 0.071 & 2.44 \\
  0.44 & 0.56 & 0.50 &  351 &  3753 & 0.09 & 0.078 & 3.24 \\
  0.56 & 0.68 & 0.60 &  286 &  3412 & 0.08 & 0.085 & 3.55 \\
  0.68 & 0.87 & 0.74 &  227 &  3871 & 0.06 & 0.097 & 3.83 \\
  0.87 & 2.00 & 1.01 &  299 &  3439 & 0.09 & 0.087 & 5.81 \\
\hline
\end{tabular}
\caption{Same as Table \ref{longEUVI} but for CR-2099.}
\label{longAIA}
\end{center}
\end{table*}

For the four smaller loop length bins in Table \ref{longEUVI}} for CR-2081, Figure \ref{phi_L} shows the statistical distributions of all energy flux quantities $\phi_r$, $\phi_c$, and $\phi_h$ for the up and down loops only, i.e. without the quasi-isothermal loops. To save space the same graph for CR-2099 is not included.

\begin{figure*} 
\begin{center}
\begin{minipage}{0.48\textwidth}
\includegraphics[width=\textwidth]{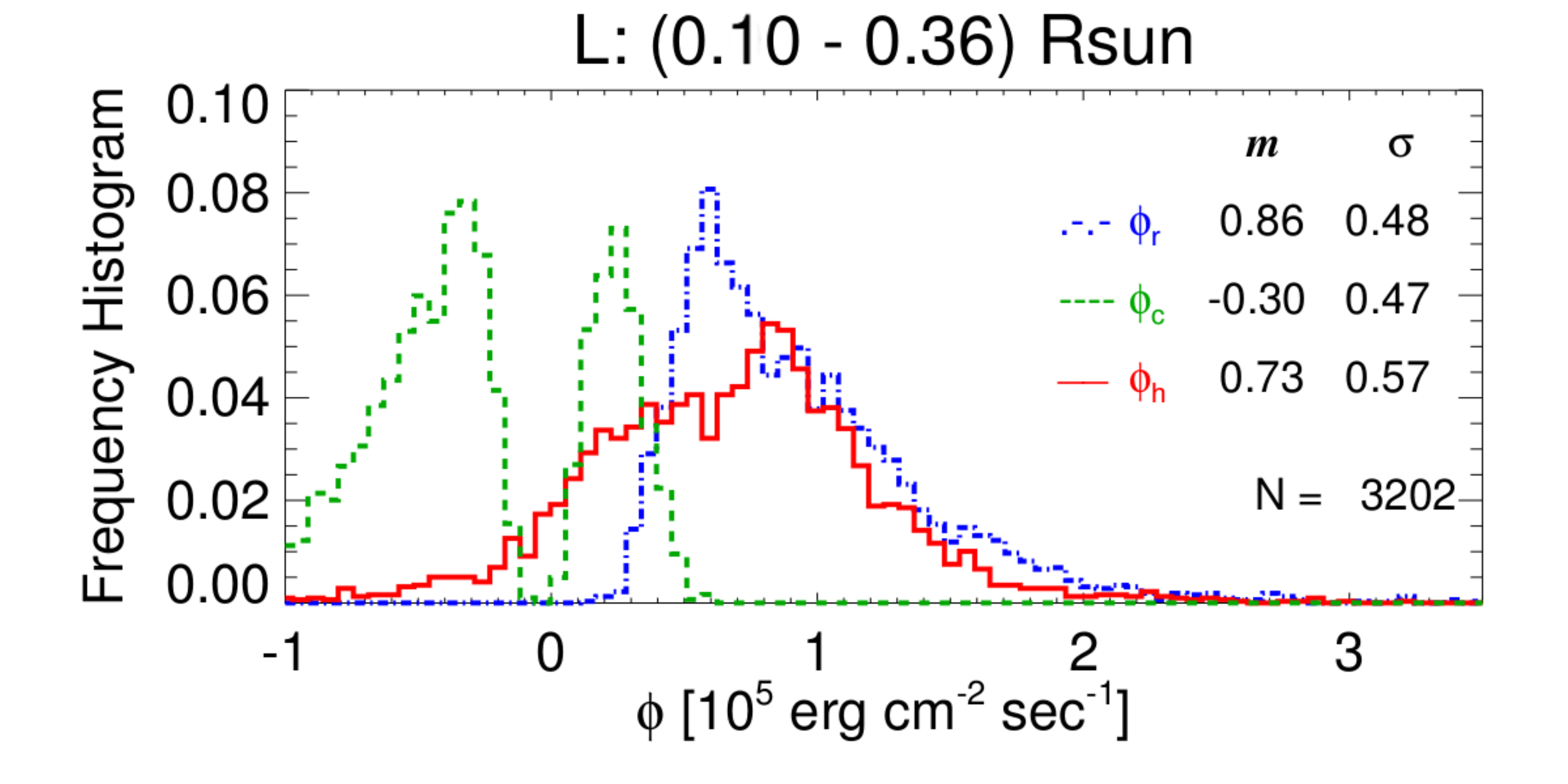}
\includegraphics[width=\textwidth]{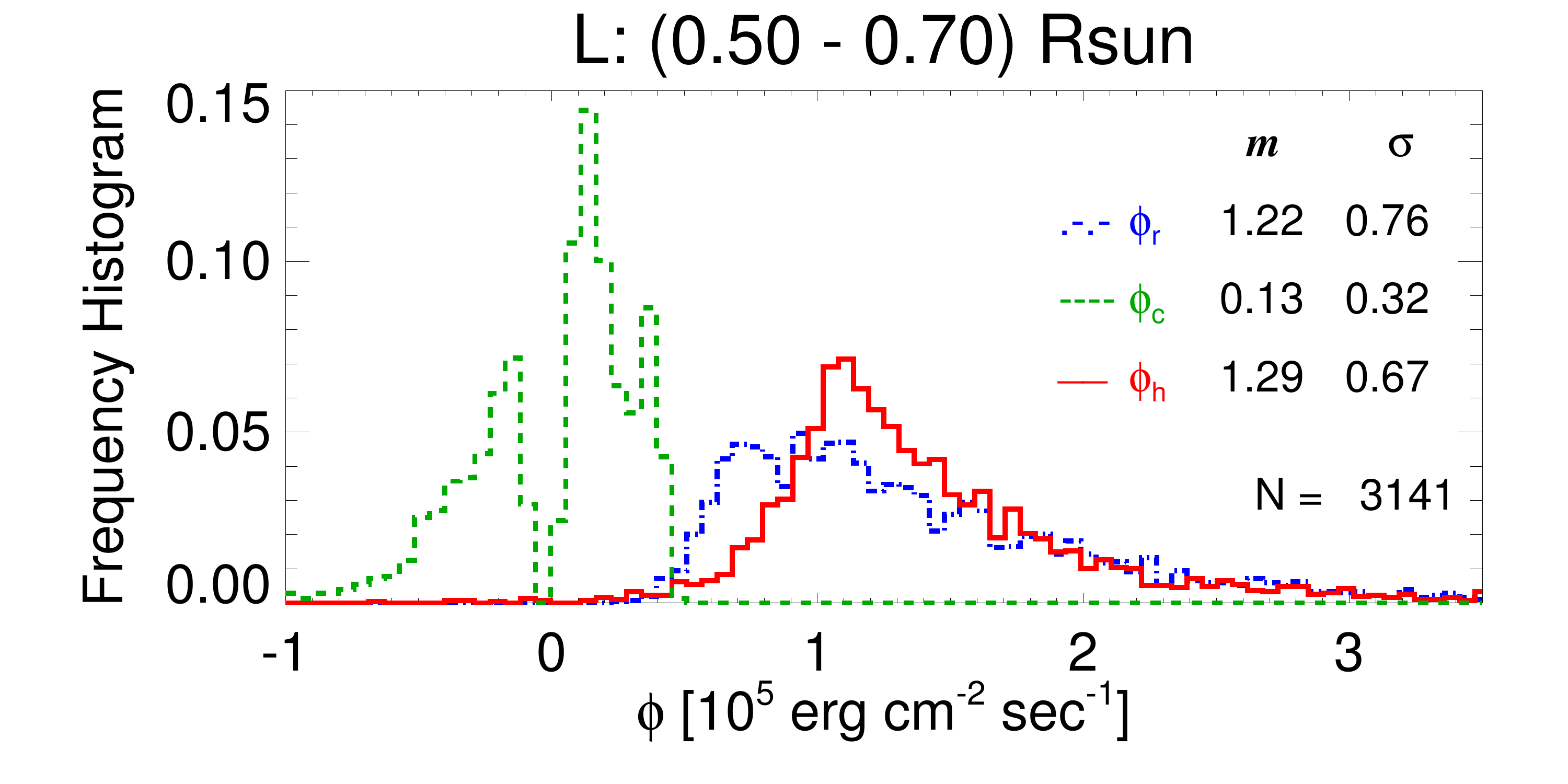}
\end{minipage}
\begin{minipage}{0.48\textwidth}
\includegraphics[width=\textwidth]{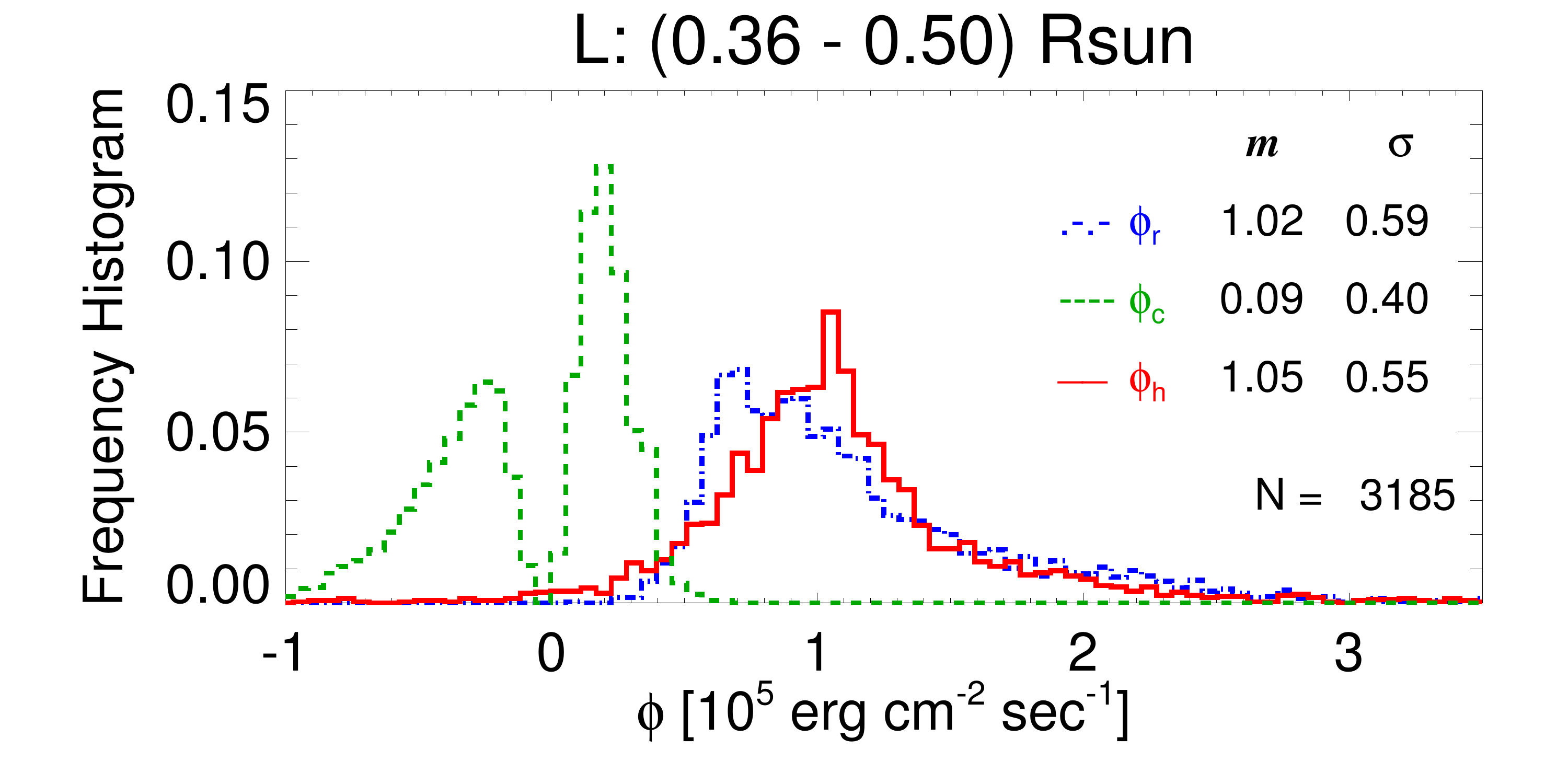}
\includegraphics[width=\textwidth]{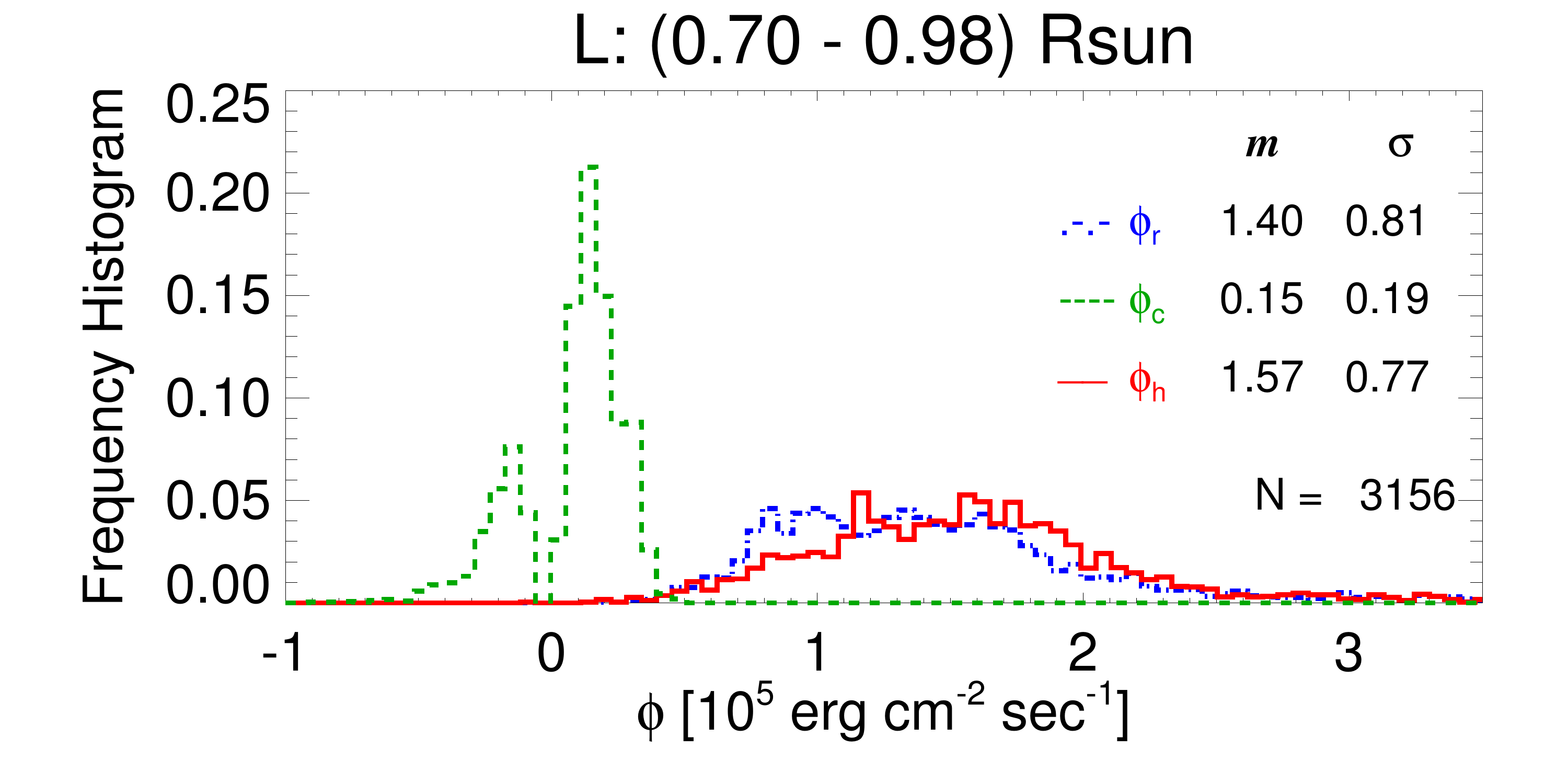}
\end{minipage}
\end{center}
\caption{Statistical distribution of the energy flux quantities $\phi_r$, $\phi_c$, and $\phi_h$ of the up and down loops for CR-2081, discriminated in the four different loop length bins defined in Figure \ref{histololos}. (A color version of this figure is available in the online journal).}
\label{phi_L}
\end{figure*}

Note that, after having filtered out the quasi-isothermal loops, all four panels of Figure \ref{phi_L} exhibit a distribution of $\phi_c$ lacking its population with values $\sim 0$, as expected. Next, note how the number of down loops, measured by the area with $\phi_c<0$, decreases with increasing mean loop length $\left<L\right>$. This is quantitatively measured in Table \ref{longEUVI} in the column $N_{\rm down}/N_{\rm tot}$. The last column in that table shows the ratio between the half length $\left<L/2\right>$ of the loops with their mean scale height $\left<\lambda_N\right>$. {Note that this ratio increases with increasing loop length, being of order $\sim 3$ in between the second and third bins, where down loops start to become less prominent (less than 50\% of the population). This is consistent with a stability criteria of down loops theoretically derived by \citet{serio_1981} that predicts down loops to be unstable for structures with values of this ratio grater than $\sim 3$.} 

Figure \ref{phi_L} shows that the integrated radiative loss quantity $\phi_r$ increases with loop length $L$. In this analysis each bin contains a mix of all types of temperature structures (up and down) and latitudes (low and mid), so the structures grouped in each panel of the Figure (corresponding to each bin in Table \ref{longEUVI}) have a similar average temperature. Consequently, the mean radiative loss is the same for all bins, and hence the characteristic value of the radiative loss quantity $\phi_r$ increases with loop length $L$ due to an increasing integral length. Consistently, the {input energy flux} $\phi_h$ also increases with loop length. In the case of CR-2099 down loops are much less prominent, and their tendency of a decreasing population with increasing loop length is much subtler, though still verified. 

\begin{table} 
\begin{center}
\begin{tabular}{|c|c|c|c|}
\hline 
   & $\left<\phi_r\right> (\sigma)$ & $\left<\phi_c\right> (\sigma)$ & $\left<\phi_h\right> (\sigma)$ \\
  \hline \hline
 up   & 0.98 (0.64) &  0.20 (0.10) & 1.21 (0.61) \\
 down & 1.27 (0.83) & -0.29 (0.31) & 0.94 (0.93) \\
 \hline
\end{tabular}
\caption{Statistical results of the quantities $\phi_r, \phi_c, \phi_h$ $[10^5\,\rm{erg\,cm^{-2}\,sec^{-1}}]$ for CR-2081{, discriminating up and down loops.}}
\label{updowncomp}
\end{center}
\end{table}

{For a very marginal population of loops, it is found that $\phi_h<0$, which is an unphysical result. This affects only the smallest loops, as revealed by the analysis of structures as a function of their size (Figure \ref{phi_L}). The radiative loss term is calculated based on plasma emission detected by the 3 coronal bands of EUVI and the 3 used of AIA. Though this should account for most of the coronal plasma, there surely is additional emission that is out of the range of sensitivity of the used instruments. Thus, the radiative loss term $\phi_r$ is most probably under-estimated which, along with the fact that this is an additive positive term in Equation (\ref{phis}), helps to explain slightly negative values of $\phi_h$. Also, the systematic errors of the DEMT technique derived by \citet{lloveras_2017}, once propagated into the energy flux quantities (which will be informed in a future work) can easily explain the marginal population of loops with negative values of $\phi_h$.}

Finally, statistical results of all the flux quantities $\phi_r$, $\phi_c$, and $\phi_h$ for CR-2081, discriminating up and down loops, are shown in Table \ref{updowncomp}. Note that $\left<\phi_r\right>$ is larger for down loops, consistently with their location at low latitudes where temperatures are lower. Down loops are also characterized by a negative conductive flux term $\left<\phi_c\right>$, which contributes then to the input of energy in the balance Equation (\ref{Balance}). Thus, a smaller value of the input energy flux $\left<\phi_h\right>$ is needed for down loops (compared to up loops) to compensate for the radiative flux and keep them stable. {As a closing comment, note that as this study deals with the coronal section of magnetic loops, the characteristic values of the conductive flux at their base are smaller than those of the radiative losses, because the temperature gradient there is not as large as in the layers underlying the Corona. Still, the conductive flux plays a part in the balance Equation (10), as shown by the histograms in Figure 8.}

\section{Statistical Comparison with a Hydrodynamic Loop Model} \label{ebtel}

{In this section, a first comparison between the results from Section \ref{phis_results} and a theoretical model is carried out, specifically using the 0D hydrodynamic model \emph{Enthalpy-Based Thermal Evolution of Loops} (EBTEL \citet{klimchuk_2008}). Here we highlight the key aspects of the comparison, and the reader is referred to \citet{maccormack_2017} for full details.} 

{EBTEL considers} the time dependent equation of energy conservation, assuming a constant area along the magnetic loop, and a piece-wise continuous radiative loss function $\Lambda_E(T_{E})$ given by \citet{klimchuk_2008}. The model separates the loop in two segments, one corresponding to the corona and the other one to the {transition region (TR). By integrating the energy balance equation in each segment, the following relationship between the radiation loss in the TR and the downwards conductive flux in the corona is obtained,}

\begin{equation}
\label{eq:ConsInt}
H_0 \approx -F_0-\phi_{r,TR}
\end{equation}

\noindent
where $H_0$ and {$F_0<0$ are the enthalpy and conductive flux at the coronal base, respectively,} and {$\phi_{r,TR}>0$} is the radiative loss flux in the TR. Two {qualitatively diverse behaviors are predicted}:

\begin{itemize}

\item If $|F_0|>\phi_{r,TR}$ {the excess conductive flux implies a positive enthalpy flux, so that matter evaporates to the corona, increasing its density.}

\item If $|F_0|<\phi_{r,TR}$ there is a deficit of conductive flux {which results in a negative enthalpy flux (promoting radiation in the TR), so that matter condensates from the corona, decreasing its density.}

\end{itemize}

{Treating the loop length and its mean coronal heating rate as free parameters,} EBTEL combines the {equations of the corona and the TR} to {predict} the temporal evolution of the pressure $\bar{P}_E$, electron temperature $\bar{T}_E$ and electron density $\bar{N}_E$ {height-averaged along the loop \citep{klimchuk_2008}. For the purpose of the comparison, the values of the two free parameters are drawn from the DEMT+PFSS results of the previous section. The average heating rate is then set equal to the input energy flux divided the loop length, $\phi_h/L$, for each magnetic field line in the model}.

The comparative analysis was performed for every loop of the sample corresponding to each latitudinal region of the two analyzed rotations in the previous section. For each loop, the ratio between the loop-height average of the DEMT electron density $\bar{N}_e$ and the EBTEL average density $\bar{N}_E$ is computed, as well as the ratio between the two average temperature, i.e. $\bar{T}_m/\bar{T}_E$. Table \ref{CompEBTEL} summarizes the median value and standard deviation of these two ratios for each of the two latitudinal regions in both analyzed rotations. 

\begin{table*}
\begin{center}
\begin{tabular}{|c|c|c|c|c|}
\hline 
    & Latitude & {$\left<\bar{N}_e/\bar{N}_{E}\right>(\sigma)$} &
    {$\left<\bar{T}_m/\bar{T}_{E}\right>(\sigma)$} \\     
  \hline \hline
CR-2081 & Low & 2.0 (1.5) & 1.1 (0.4) \\
 (EUVI) & Middle & 2.3 (1.3) & 1.1 (0.2) \\
  \hline \hline
CR-2099 & Low & 2.2 (1.4) & 1.1 (0.3) \\
 (AIA)  & Middle & 2.4 (1.1) & 1.0 (0.2) \\
\hline
\end{tabular}
\caption{{Statistics of the ratio between, a) the loop-height averaged DEMT electron density $\bar{N}_e$ and the EBTEL average electron density $\bar{N}_{E}$, and b) the loop-height averaged DEMT electron temperature $\bar{T}_m$ and the mean EBTEL electron temperature $\bar{T}_{E}$. Values corresponding to low and mid-latitudes for CR-2081 and CR-2099 are discriminated.}}
\label{CompEBTEL}
\end{center}
\end{table*}

While the DEMT and EBTEL electron temperatures are similar for all analyzed populations, the electron densities differ by an average factor of $\sim 2.2$. This systematic discrepancy can be traced down to assumptions in the EBTEL model. Most importantly, specific physical scale laws that are used in the model, as well as the characteristic size assumed for the modeled loops (of the order of their thermal scale height), are consistent with observations of AR loops rather than with quiet-Sun coronal structures. Furthermore, the radiation loss function used by EBTEL and the one used in Section \ref{phis_results} differ in the temperature range of interest, which has an impact on the derived densities. All these factors build up to explain the observed systematic difference in densities \citep{maccormack_2017}.

\section{Discussion and Conclusions} \label{conc}

A new DEMT tool was developed that allows calculation of the energy input flux $\phi_h$ required at the coronal base ($r\sim\,1.025\,{\rm R}_\odot$) of magnetic loops of the quiet-Sun corona to sustain hydrostatic thermodynamically stable structures. First results of applying the tool to two solar rotations (CR-2081 and CR-2099) with different level of activity are shown. The characteristic values obtained are in the range $\phi_h\sim\,0.5-2.0\times 10^5\,{\rm (erg\,sec^{-1}\,{\rm cm}^{-2})}$, depending on the particular coronal structure and the level of activity of the corona.

For CR-2081, a solar minimum rotation, the mid-latitude hotter regions of the streamer belt and the cooler low-latitude regions exhibit similar median values in their distribution of energy input flux, {with mid-latitudes characterized by} a considerably smaller standard deviation. The same characteristics are observed for CR-2099.

For CR-2099, during the early rising phase of SC 24, the analysis was performed with both the EUVI and AIA instruments. Results obtained with both instruments are highly consistent in the mid-latitude region, where the sample size is similar, with a mean value of the energy input flux in this region being about $\sim 20\%$ larger than during solar minimum. In the low latitudes, the results with both instruments are somewhat less consistent (though still comparable), being the case that the AIA data produced a considerably larger population size. The low-latitude results of CR-2099, based in AIA, show virtually the same energy input fluxes as the the low latitudes of CR-2081, based on EUVI.

The characteristic values of energy input flux in different sub-regions of the equatorial streamer belt are related to the presence of different types of thermodynamic structures, namely the up and down loops first discovered by \citet{huang_2012} and further studied by \citet{nuevo_2013}. The study here presented added new insight on the characteristics of down loops, showing that they are characterized by smaller values of energy input flux due to the extra energy source of heat conduction, {and that their population is larger for smaller scales (see Table \ref{longEUVI}).}

The characteristic values obtained for the energy input flux $\phi_h$ in the quiet-Sun corona are consistent with observational estimates for the quiescent corona reported in previous works by \citet{withbroe_1977}, \citet{aschwanden_2004}, and more recently by \citet{hahn_2014}. Based on spectroscopic data of quiet sun regions taken by the Extreme Ultraviolet Imaging Spectrometer (EIS) at a height range $1.05-1.20\,{\rm R}_\odot$, on board the \emph{Hinode} mission, \citet{hahn_2014} estimate the non-thermal component in the observed broadening of spectral lines. Assigning the non-thermal line broadening to Alfvén waves, they derive the corresponding wave energy flux as a function of height, estimating the local plasma density based on DEM analysis, and the local Alfvén speed relying on a magnetic potential extrapolation to estimate the magnetic field strength. Their study includes observations both around the equator and mid-latitudes, for which the authors find their respective distributions of wave energy flux at the coronal base to be in the range $\sim (2.0\pm 0.4)$ and $\sim (1.8\pm 0.5) \times 10^5\,{\rm (erg\,sec^{-1}\,{\rm cm}^{-2})}$, respectively, which compare well to the characteristic distributions found in this work (Figures \ref{flujo2081EUVI} and \ref{flujo2099AIA}). Based on those estimates, a large fraction of the coronal base energy input flux $\phi_h$ estimated in this work, or even its totality, could be accounted for by Alfvén waves.

Under the assumption that all the energy input flux is in the form of Alfvén waves, their quadratic velocity amplitude $\left<\delta v^2\right>$ is related to the input energy flux through $\phi_h = \rho\, \left<\delta v^2\right>\,V_A$ \citep{moran_2001,hahn_2014}, where $\rho$ and $V_A = B/\sqrt{4\pi\,\rho}$ are the local plasma mass density and Alfvén speed. Accounting for a $\sim 8\%$ helium abundance in the corona, the mass density can be estimated as $\rho\approx 1.14\,m_p\,N_e$ in terms of the electron density $N_e$ and the proton mass $m_p$. Applying this relationship to each magnetic field line, using the PFSS and DEMT models to estimate $B$ and $N_e$, respectively, at the coronal base, the range obtained for the energy input flux translates into a characteristic Alfén wave velocity amplitude range $\sqrt{ \left<\delta v^2\right>}\sim\,25-40\,{\rm (km/seg)}$. The upper limit corresponds to the low-latitudes in the streamer belt of CR-2081, and the lower limit to mid-latitudes. This range of Alfvén wave amplitudes, consistent with the quiet-Sun coronal estimates by \citet{hahn_2014}, are also in agreement with characteristic ranges reported in coronal hole studies, see for example \citet{banerjee_2011} and references therein.

Based on a 0D HD physical model for coronal loops \citep{klimchuk_2008}, we confirmed that the characteristic values obtained for the coronal base energy input flux $\phi_h$ are consistent with the height-averaged values of electron temperature and density obtained from the DEMT analysis. In a future effort, results will be compared to 1D HD models.

In a recent work, \citet{nuevo_2015} extended the DEMT technique to also use the 335 \AA\, coronal band of the AIA instrument, which, combined with the other 3 bands used in this paper, expands the sensitivity range to 0.5-4.0 MK. In that work, the authors find a ubiquitous bimodal coronal LDEM distribution, with two distinct characteristic temperatures. Their result is interpreted as revealing the ubiquitous presence of ``warm" ($T\sim 1.5\,{\rm MK}$) and ``hot" ($T\sim 2.6\,{\rm MK}$) loops throughout the quiet-Sun closed corona. The loops analyzed in this paper correspond to the warm class. In a future work this new tool will be applied to also study the hot loops.

The new tomographic tool developed in this paper provides a semi-empirical constraint to global coronal heating models. Its main new product is in the form of 2D maps of the energy input flux $\phi_h$ at the coronal base layer of the quiet-Sun closed corona. This will be used as a validation tool at the coronal base layer of {steady-state 3D MHD coronal simulations of} the Space Weather Modeling Framework (SWMF), developed by \citet{vanderholst_2014}. This will be the subject of a future effort. 

{DEMT studies provide a time-averaged description of the state of the corona during the observing time of each structure ($\sim 1/2$ solar rotation). The new results of this study are to be interpreted in this context, and can be compared with steady-state models of large-scale coronal structures, such as those recently developed by \citet{schiff_2016}, who sucessfully reproduced the up/down loops reported by \citet{huang_2012,nuevo_2013}, and studied here in Section \ref{updown}.}

\acknowledgments 

{The authors wish to thank the anonymous referee for a careful reading of the manuscript and insightful review, which helped to improve the content and clarity of the manuscript.} C.MC. acknowledges CONICET doctoral fellowship (Res. Nro. 4870) to IAFE that supported her participation in this research. C.MC., A.M.V., F.A.N., and M.L.F. acknowledge ANPCyT grant PICT \# 2012/0973 and CONICET grant PIP \# 11220120100403 to IAFE that partially supported
their participation in this research.

\end{document}